\documentclass[preprint2]{aastex}
\shorttitle{{\sl HST}/ACS Observations of I~Zw~18: 
a Young Galaxy in Formation}
\shortauthors{Izotov \& Thuan}
\begin{document}

\title{Deep {\sl Hubble Space Telescope}/ACS 
Observations of I~Zw~18: 
a Young Galaxy in Formation\footnote{Based on observations 
obtained with the NASA/ESA {\sl Hubble Space Telescope} through the Space 
Telescope Science Institute, which is operated by AURA,Inc. under NASA
contract NAS5-26555.}}

\author{Yuri I. Izotov}
\affil{Main Astronomical Observatory, Ukrainian National Academy of Sciences,
27 Zabolotnoho, Kyiv 03680, Ukraine}
\email{izotov@mao.kiev.ua}

\and

\author{Trinh X. Thuan}
\affil{Astronomy Department, University of Virginia, Charlottesville, VA 22903, USA}
\email{txt@virginia.edu}

\begin{abstract}
We present $V$ and $I$ photometry of the resolved stars in the 
most metal-deficient blue compact dwarf galaxy known, I~Zw~18 ($Z_\odot$/50), 
using {\sl Hubble Space Telescope}/Advanced Camera for Surveys (ACS)
images, the deepest ones ever obtained for this galaxy. 
The resulting $I$ vs. $V-I$ color-magnitude diagram (CMD) reaches 
limiting magnitudes $V$ = $I$ = 29 mag. 
It reveals a young stellar population of blue main-sequence (MS) stars
(age $\la$ 30 Myr) 
and blue and red supergiants (10 Myr $\la$ age $\la$ 100 Myr), but also an 
older evolved population of
asymptotic giant branch (AGB) stars (100 Myr $\la$ age $\la$ 500 Myr). 
We derive a
distance to I~Zw~18 in the range 12.6 Mpc -- 15 Mpc 
from the brightness of its AGB stars, with preferred values in the higher 
range.
The red giant branch (RGB) stars are conspicuous by their absence,
although, for a distance of I Zw 18 $\leq$ 15 Mpc, 
our imaging data go $\sim$ 1 -- 2 mag below the tip of the RGB.
Thus, the most evolved stars in the galaxy are not older than 500 Myr and 
{\it I~Zw~18 is a bona fide young galaxy}. 
Several star formation episodes can be inferred from the CMDs 
of the main body and the C component.
There have been respectively three and two episodes in these two parts, 
separated by periods of $\sim$ 100 -- 200 Myr.
In the main body, the younger 
MS and massive post-MS stars are distributed over a
larger area than the older AGB stars, suggesting that I~Zw~18 is still
forming from the inside out. In the C component,
different star formation episodes are spatially distinct, with 
stellar population ages decreasing from the northwest to the southeast,
also suggesting the ongoing build-up of a young galaxy.
\end{abstract}


\section{Introduction}

The question of whether there are young galaxies in the local 
universe forming stars for the first time  
is of considerable interest for galaxy formation and cosmological 
studies. There are several reasons for this.
First, Cold Dark Matter models predict that low-mass dwarf galaxies 
could still be forming at the present epoch because they originate 
from density fluctuations considerably smaller than those 
giving rise to the giant ones. Thus the existence of young dwarf galaxies 
in the local universe would put strong constraints on the 
primordial density fluctuation spectrum. 
Second, while much progress has been made in 
finding large populations of galaxies at high ($z$ $\geq$ 3) redshifts 
\citep[e.g.][]{S96}, truly young galaxies in the 
process of forming remain elusive in the distant universe. 
The spectra of those far-away galaxies 
generally indicate the presence of a substantial amount of heavy elements, 
implying previous star formation and metal enrichment. Thus it is important
to have examples of bona fide young galaxies in the local universe because 
they can be used as laboratories to study star formation and 
chemical enrichment processes in 
environments that are sometimes much more pristine than those in known 
high-redshift galaxies. Moreover, their proximity allows studies of their 
structure, metal content, and stellar populations with a sensitivity, 
precision, and spatial resolution that faint distant high-redshift galaxies 
do not allow. Finally, in the hierarchical model of galaxy formation,
large galaxies result from the merging of smaller structures. 
These building-block galaxies are too faint and small to be studied at 
high redshifts, while we stand a much better chance of understanding them 
if we can find local examples.

The blue compact dwarf (BCD) galaxy I~Zw~18 is one of the best candidates 
for being a truly young galaxy.
\citet{Z66} described it as a double system of compact galaxies,
which are in fact two bright knots of star formation with an angular 
separation of 5\farcs8. 
These two star-forming regions are referred to as the 
brighter northwest (NW) and fainter southeast (SE) components.
They form what we will refer to as the main body (Figure \ref{Fig1}). 
The presence of ionized gas emission \citep[e.g.][]{HT95} and of Wolf-Rayet
stars in the NW component \citep{I97,L97,B02} suggest active ongoing star formation
in the main body.
Later studies by \citet{D89} and \citet{DH90} have revealed a more complex 
optical morphology.
They pointed out the existence of a
prominent diffuse feature $\sim$ 22\arcsec\ northwest of the 
main body, hereafter called the C component 
(Figure \ref{Fig1}), which contains a blue irregular star-forming region. 
\citet{D96a}, \citet{IT98} and \citet{V98} have shown the C component
to have a systemic radial velocity equal to that of the ionized gas in the 
main body, thus establishing its physical association to I\ Zw\ 18. 
Furthermore, the interferometric H {\sc i} map of I~Zw~18 by
\citet{V98} has shown that the C component is
embedded in a common H {\sc i} envelope with the main body. 

I~Zw~18 remains the most metal-poor BCD and the
lowest metallicity star-forming galaxy known
since its discovery by \citet{SS70}.
Later spectroscopic observations by \citet{SS72}, \citet{L79}, \citet{F80},
\citet{KD81}, \citet{P92}, \citet{SK93}, \citet{IT98} and \citet{I99}
have confirmed its low metallicity, with an oxygen abundance of only 
$\sim$ 1/50 the solar value.

Many studies have focused on the evolutionary state of
I~Zw~18. \citet{SS72} and \citet{HT95}
have suggested that it may be a young galaxy,
recently undergoing its first burst of star formation. The latter authors 
concluded from {\sl Hubble Space Telescope} ({\sl HST})/Wide Field and 
Planetary Camera 2 (WFPC2) images that the colors of 
the diffuse unresolved component surrounding the SE and NW  regions are 
consistent with a population of B and early A stars, with no evidence for 
older stars. 
From color-magnitude diagram (CMD) studies of the main body based on other
{\sl HST}/WFPC2 images, 
\cite{D96b} have found that star formation 
in I~Zw~18 began at least 30 -- 50 Myr ago and is still
continuing to the present. 
\cite{A99} have discussed the star formation
history in I~Zw~18 using the same {\sl HST}/WFPC2 data as
\citet{HT95} and \citet{D96b}, but going deeper thanks to a more
sophisticated treatment of the data. 
By comparing the observed CMDs and luminosity functions with the
synthetic ones, they 
concluded that there were two episodes of star formation in I~Zw~18.
The first star formation episode occurred $\sim$ 0.5 -- 1 Gyr ago, as 
evidenced by the presence of asymptotic giant branch (AGB) stars.
This age is more than 10 
times larger than the one obtained by \citet{D96b}. The more intense 
second star formation episode took place between 15 and 20 Myr ago.
Subsequently, \citet{O00} has carried out a near-infrared (NIR) 
CMD study based on {\sl HST}/Near Infrared Camera and Multi Object 
Spectrometer (NICMOS) $J$ (F110W) 
and $H$ (F160W) images. He concluded that 
the main body of I~Zw~18 is dominated by 
two populations, a 10 -- 20 Myr population of red supergiants (RSG)
and a considerably older (0.1 -- 5 Gyr) population of AGB stars.

Because of its faintness, the C component has not been studied in as much 
detail. Its optical 
spectrum \citep{IT98,I01,V98} reveals a blue continuum with weak Balmer 
absorption features and faint H$\alpha$ and H$\beta$ in emission,
suggesting that the H {\sc ii} region is ionized 
by a population of early B stars.
In a $B-V$ vs. $V$ CMD analysis 
of the C component, \citet{D96b} have found  a well-defined 
stellar upper main-sequence (MS) of blue stars with ages $\sim$ 40 Myr. 
However, numerous faint red stars are also present,
with ages $\sim$ 100 -- 300 Myr. Those authors concluded that the C component
consists of an older stellar population with an age of several hundred Myr, 
but that it has experienced more recently a modest starburst in its 
southeastern half,
as evidenced by the presence of blue stars and H$\alpha$ emission. 
\citet{A99} have estimated an age for the C component not exceeding 200 Myr. 

Using integrated optical and NIR photometric colors derived from
ground-based and archival
{\sl HST} images, \citet{P02} and \citet{H03} have shown that the age of 
I~Zw~18 as a whole, including both
the main body and the C component, does not exceed 500 Myr. 
This is in agreement with the upper age limits obtained by \citet{HT95}, 
\citet{A99} and \citet{D96b}, but in contradiction with the large one
of 5 Gyr obtained by \citet{O00}.

\citet{L00a} and \citet{L00b} have proposed an extreme scenario in which
star formation in I~Zw~18 proceeds, not only in 
recent short episodic bursts, but also in 
a continuous fashion at a low level over a long period of many Gyr.
This would  
result in an extended underlying low-surface-brightness (LSB) red stellar 
component in I~Zw~18. However, \citet{I01} 
and \cite{P02} have shown that the 
existence of such a component is supported neither by spectroscopic nor 
photometric measurements of the extended emission around I~Zw~18. 

It is evident that despite considerable work, the debate on the age of 
I~Zw~18 is still not settled. The evolutionary status of I~Zw~18 remains 
elusive because of several limitations in the preceding work. Previous 
studies of stellar populations in I~Zw~18 based on integrated photometric
colors or on spectral synthesis do not provide a unique star formation
history. CMD studies constrain better the star formation history, but
until now the available {\sl HST}/WFPC2 images on which these studies are
based did not go deep enough below the tip of the red giant branch (TRGB)
to reveal whether a faint old ($\ga$ 1 -- 2 Gyr) stellar population of 
red giant branch (RGB)
stars exists in I~Zw~18 or not. Evidently deeper {\sl HST} observations
were needed. The installation of the Advanced Camera for Surveys (ACS)
on {\sl HST} in 2003 made those deeper observations possible.

We use here new deep {\sl HST}/ACS imaging of I~Zw~18 in $V$ and $I$ 
to go $\sim$ 1 -- 2 mag below the TRGB and put more stringent constraints 
on the state of evolution of the BCD.
The observations and data reduction are described in Sect. 2. 
The lack of a RGB population in the CMD of I~Zw~18 
is examined in Sect. 3.
The distance determination to I~Zw~18 is discussed in Sect. 4.
CMDs of the stellar populations are studied in Sect. 5.
We find that I~Zw~18 does not contain stars older than $\sim$ 500 Myr, making
it a truly young galaxy.
Our results are summarized in Sect. 6.

\section{Observations and data reduction}

  We have obtained {\sl HST} images of I~Zw~18 
during the 2003 May 26 -- June 6 period in the course of 
eight visits totaling 25 orbits, with the ACS/Wide Field Camera (WFC) 
detector through filters F555W and F814W, 
which we will refer to hereafter as $V$ and $I$. The $V$ exposure was obtained 
during five visits totaling 16 orbits and the $I$ exposure during three 
visits with a total of 9 orbits. All observations were obtained with the same
orientation of the field of view, with one exposure per orbit,
split into two subexposures to permit cosmic ray removal. 
Drizzling was applied so that separate exposures obtained during one visit
were slightly offset with respect to one another to permit better 
spatial sampling. 
The total exposure time was 43520s ($\sim$ 12.1 hours) in $V$ and 
24300s ($\sim$ 6.8 hours) in $I$. 
The galaxy was positioned on the WFC1 frame, with the scale being 
0\farcs05 per pixel.

   Preliminary processing of the raw images 
was done at the Space Telescope Science Institute through the 
standard pipeline. This resulted in distortion-corrected and
drizzled images, six in the $V$ band and three in the $I$ band,
which we used in our 
subsequent reductions with IRAF{\footnote{IRAF is the Image Reduction and 
Analysis Facility
distributed by the National Optical Astronomy Observatory, which is
operated by the Association of Universities for Research in Astronomy
(AURA) under cooperative agreement with the National Science Foundation
(NSF).}}.
The separate $V$ and $I$ images were then combined 
with a coregistration better than 0.1 pixel and cosmic rays removed.

\begin{figure*}
\epsscale{2.2}
\plottwo{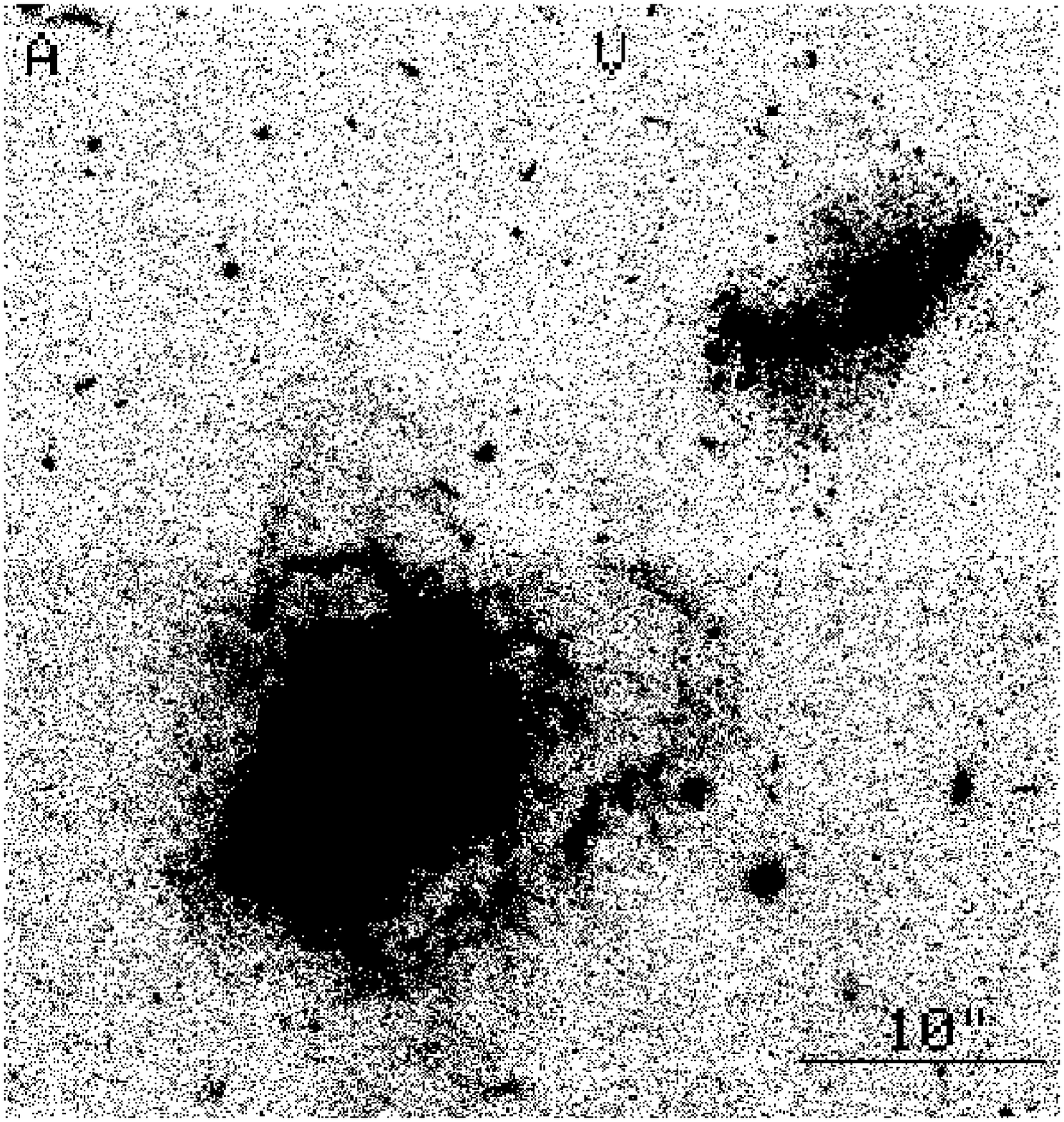}{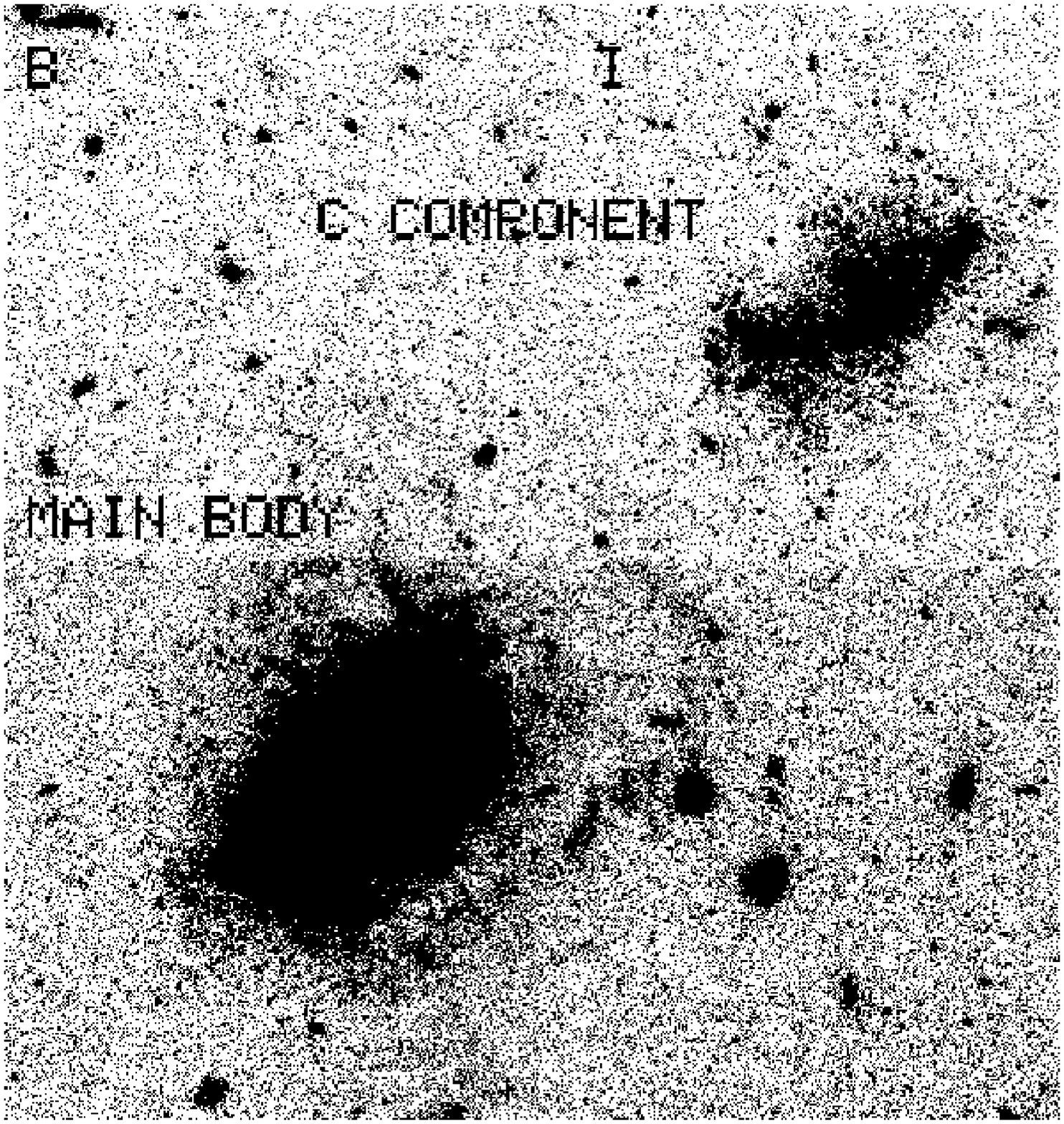}
\caption{$V$ (a) and $I$ (b) ACS images of I~Zw~18. North is up and East
is to the left. The contrast has been adjusted to show the features with
lowest surface-brightnesses. Large supershells of ionized gas can be seen
delineating supernova cavities in both the main body and the C component. 
But no extended low-surface-brightness underlying component of red 
old stars is present. The scale is shown in panel a).
\label{Fig1}}
\end{figure*}

\begin{figure*}
\epsscale{1.0}
\plotone{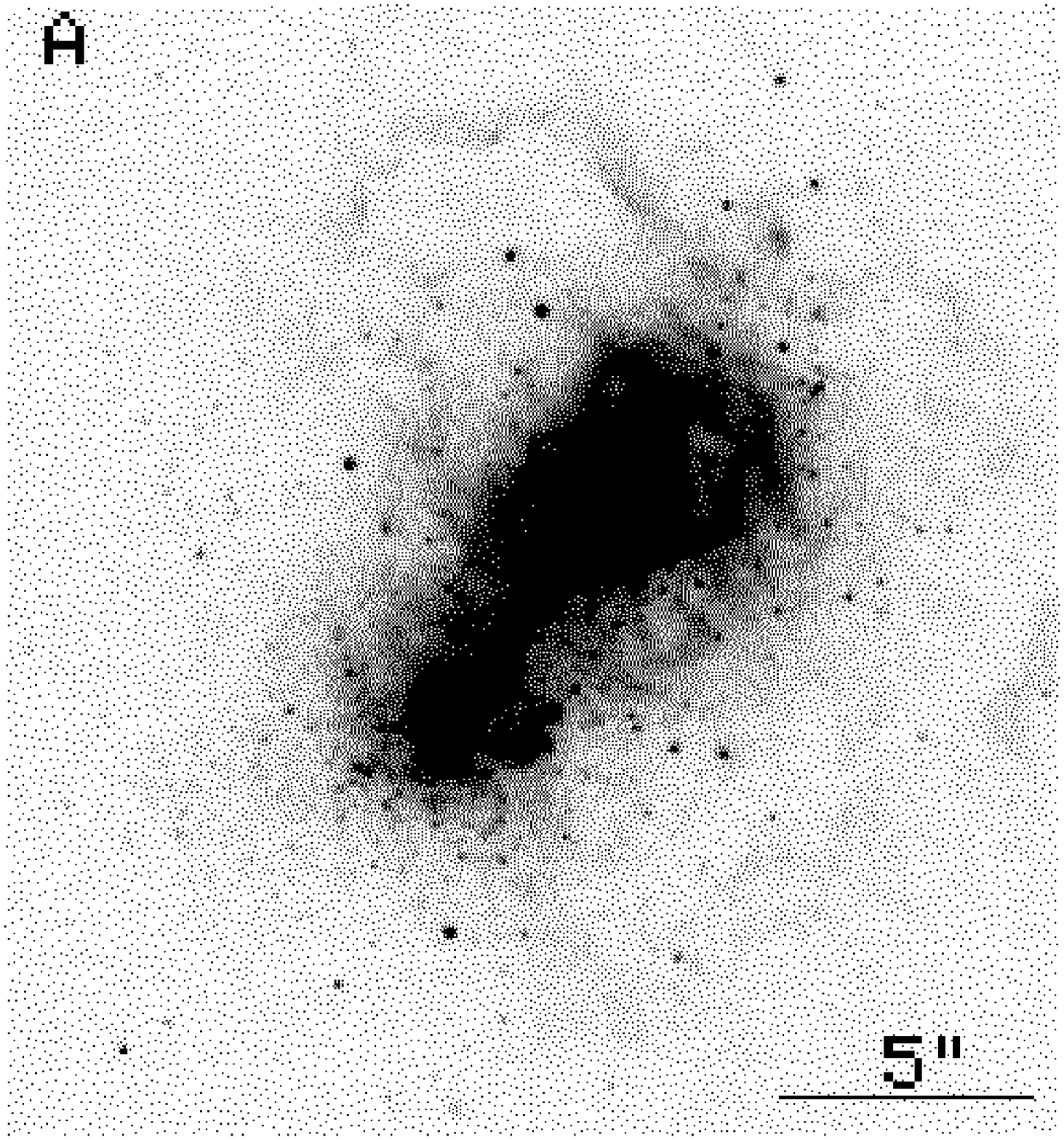}
\plotone{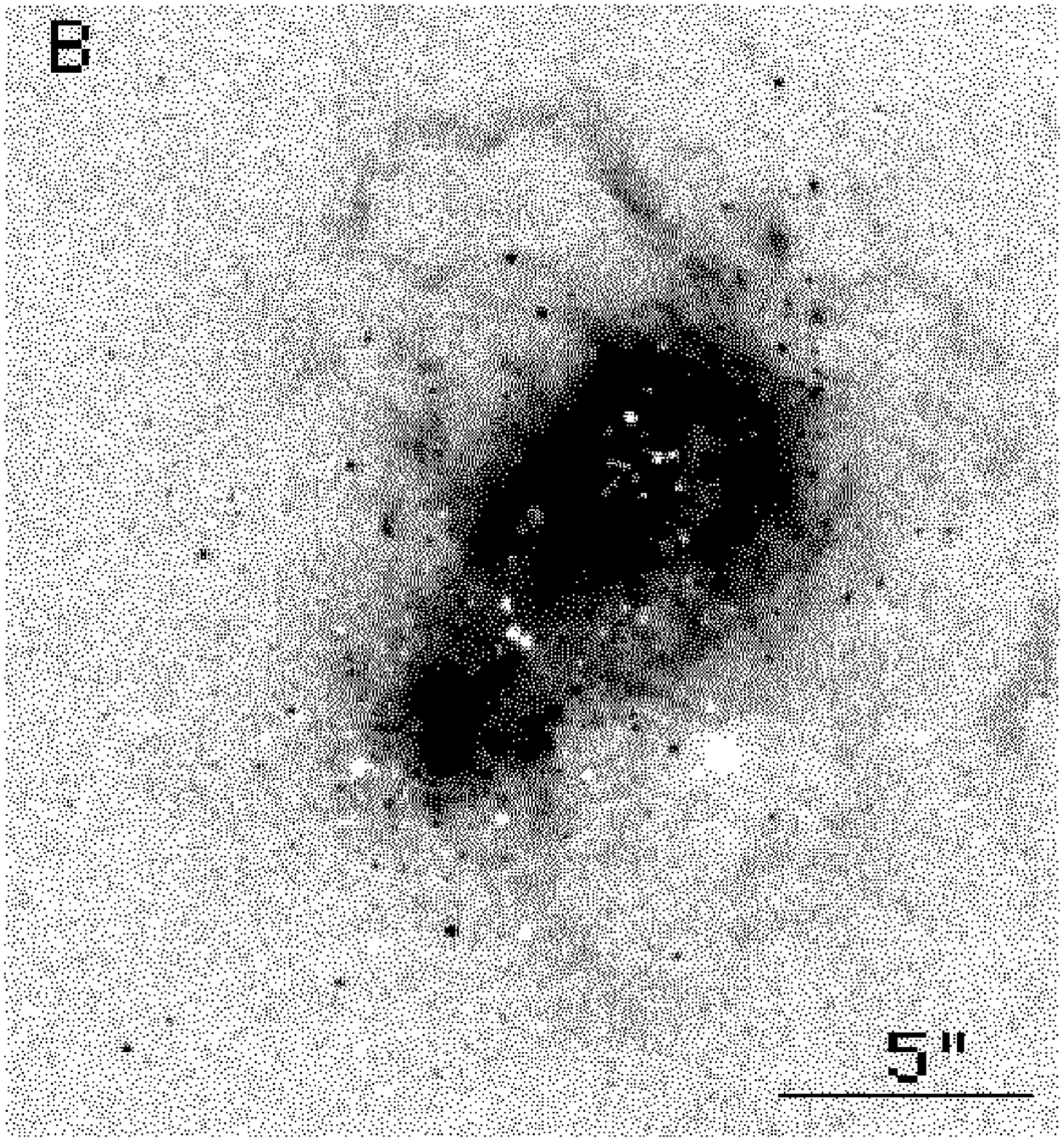}
\plotone{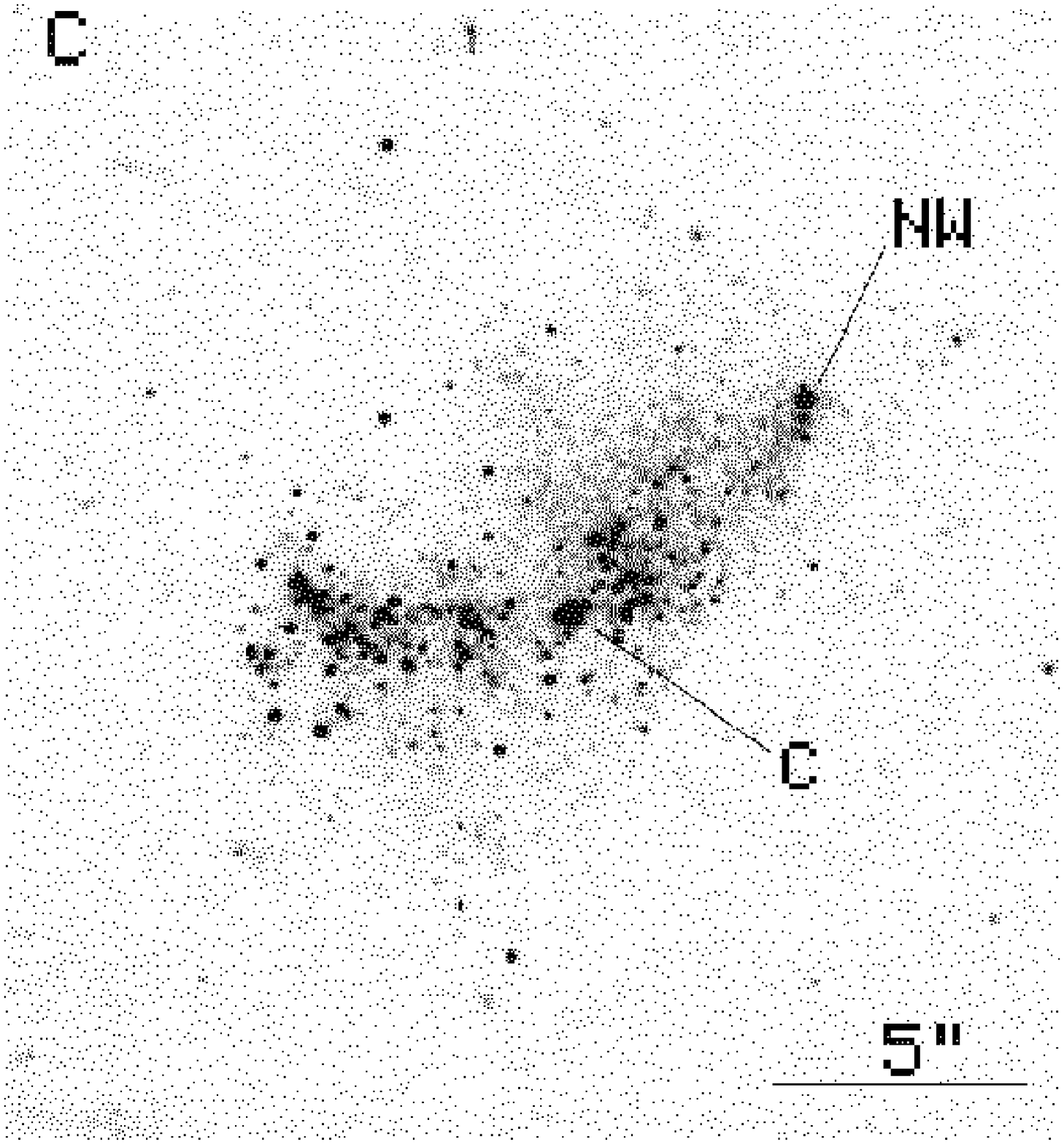}
\plotone{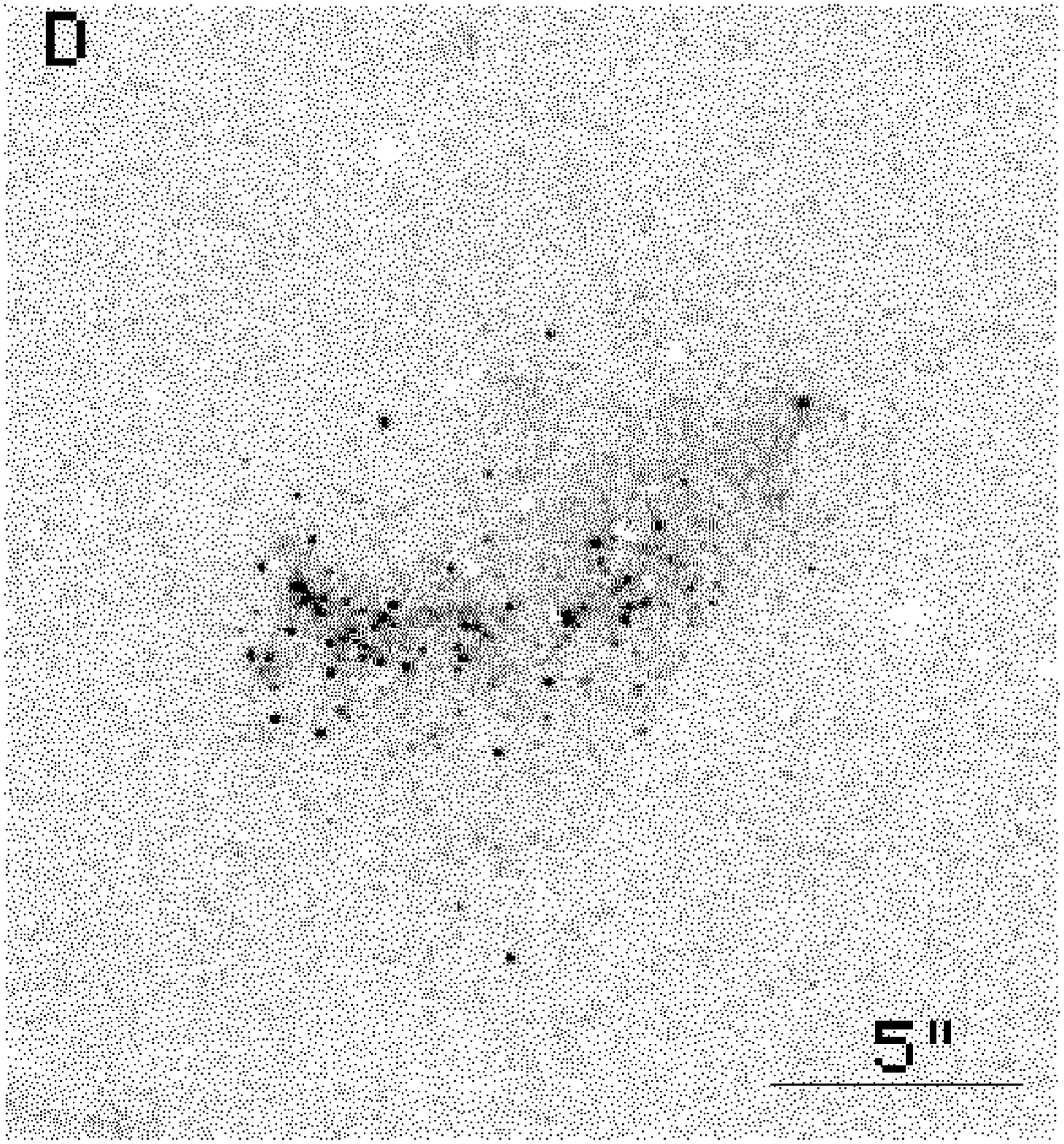}
\caption{$V$ (a) and $V-I$ (b) images of the main body. In (b) 
blue colors are dark and red colors are white. Numerous bright blue
stars can be seen scattered around the main body. Stellar emission is strongly 
contaminated by the light of the filamentary structures of ionized gas.
The extended white knot in (b) south of the NW component is a background 
galaxy. $V$ (c) and $V-I$ (d) images of the C component. 
The emission is dominated by blue stars, while the contribution of ionized 
gas emission is small. There are two stellar clusters in the C component
labeled ``C'' and ``NW'' in (c).
North is up and East is to the left. The scale is shown by a
horizontal bar in each panel.
\label{Fig2}}
\end{figure*}

 The ACS $V$ and $I$ images of I~Zw~18 are shown in 
Figures \ref{Fig1}a -- \ref{Fig1}b, rotated so that
North is up and East is to the left. The main body and the
C component are labeled in Figure \ref{Fig1}b.
Their enlarged $V$ and $V-I$ views are displayed respectively
in Figures \ref{Fig2}a -- \ref{Fig2}b and Figures \ref{Fig2}c -- \ref{Fig2}d.
The dark regions in the $V-I$ images correspond to blue colors and the white 
regions to red colors. The extended and filamentary dark regions in the 
main body 
(Figure \ref {Fig2}b) represent ionized gas emission from supernova shells, 
while extended white sources are background red galaxies. Note the presence
of several blue stars at large distances from the main body, 
suggesting recent or ongoing star formation in the halo. However, the extended
halo of red stars, which is a common feature of many dwarf galaxies 
observed with the {\sl HST} \citep[e.g. in NGC 2366, ][]{TI04}, is 
conspicuous by its absence in I~Zw~18. No appreciable ionized gas 
emission is seen in the C component (Figures \ref{Fig2}c -- \ref{Fig2}d)
despite the presence of many blue stars.

   The superior spatial resolution of the {\sl HST}/ACS images
permits to resolve individual stars and study 
stellar populations in I~Zw~18 by means of CMDs. Thanks to the higher 
sensitivity
of the ACS and longer exposure times, our data is considerably deeper than
previous {\sl HST} observations of I~Zw~18 with the WFPC2.
We used the DAOPHOT 
package in IRAF to perform point-spread-function (PSF) fitting photometry
in the sky area shown in Figure \ref{Fig1}. Thanks to the drizzling procedure,
the several bright and relatively isolated stars 
around the main body and the C component (Figures \ref{Fig2}a and
\ref{Fig2}c) allow us to derive
reliable and well sampled stellar PSFs.
We obtain full widths at half maximum (FWHM) of the stellar profiles
of 2.5 pixels (0\farcs125) and 2.8 pixels (0\farcs140) respectively 
for the $V$ and $I$ images.
As for the aperture radius to use for stellar photometry, we experimented with 
radii between
2 and 4 pixels to search for the best value giving at the same time a
high recovery rate of point sources and small
aperture corrections to the derived magnitudes.
We find that the best compromise is an aperture radius of 3 pixel
(0\farcs15). With
smaller radii the aperture corrections are too large 
(the calibrating aperture radius is 0\farcs5) and introduce unwanted 
uncertainties in the photometry. With larger radii the number of 
recovered stars is sharply reduced, because of the crowding effect which is
especially important in the main body. The background level was measured 
in an annulus with radii 4 and 6 pixels (0\farcs2 and 0\farcs3) around each 
source and subtracted. 
The zero-points of the photometry are defined as the magnitude of a star
which produces a count rate of one electron per second in a given
filter and are set so that the star Vega has magnitude 0 at all wavelengths.
This gives zero-points of 25.711 mag and 25.487 mag for $V$ (F555W filter) and 
$I$ (F814W filter)  
respectively\footnote{See the Web page at the address 
http://www.stsci.edu/hst/acs/analysis/zeropoints.}. 

Photometric aperture corrections were obtained by a comparison of
PSF-fitted magnitudes of 12 bright isolated 
stars with the magnitudes of the same stars
measured with the aperture photometry technique within a 10 pixel 
(0\farcs5) aperture. We obtained the corrections $V_{\rm ap}$(0\farcs5) -- 
$V_{\rm fit}$ = --0.18 mag and $I_{\rm ap}$(0\farcs5) -- $I_{\rm fit}$ 
= --0.23 mag. Correction for charge-transfer efficiency loss has been carried
out according to the prescriptions of \citet{R03}. We find however, that this
effect is negligible (less than 0.01 mag for sources brighter than 29 mag) 
given the high level of the night sky 
in our long-exposure images.

To check our photometry, we compare the magnitudes of several bright isolated stars in the main body, measured through a large aperture of 10 pixel radius, 
with those measured using 
the same aperture in the WFPC2 images of \citet{HT95}. We find that our 
magnitudes are
fainter by 0.05 mag in $V$ and $I$.
To put our results on the same scale as
the WFPC2 data, we have applied that small correction to our data.
Since no photometric correction to infinite aperture is yet available
for the ACS data, we have not considered it. In total, taking into account
the aperture corrections and the slight shift between the ACS and WFPC2
magnitude scales, the corrections to
our PSF photometry amount to --0.23 mag in $V$ and --0.28 mag in $I$.

\begin{figure*}
\plotfiddle{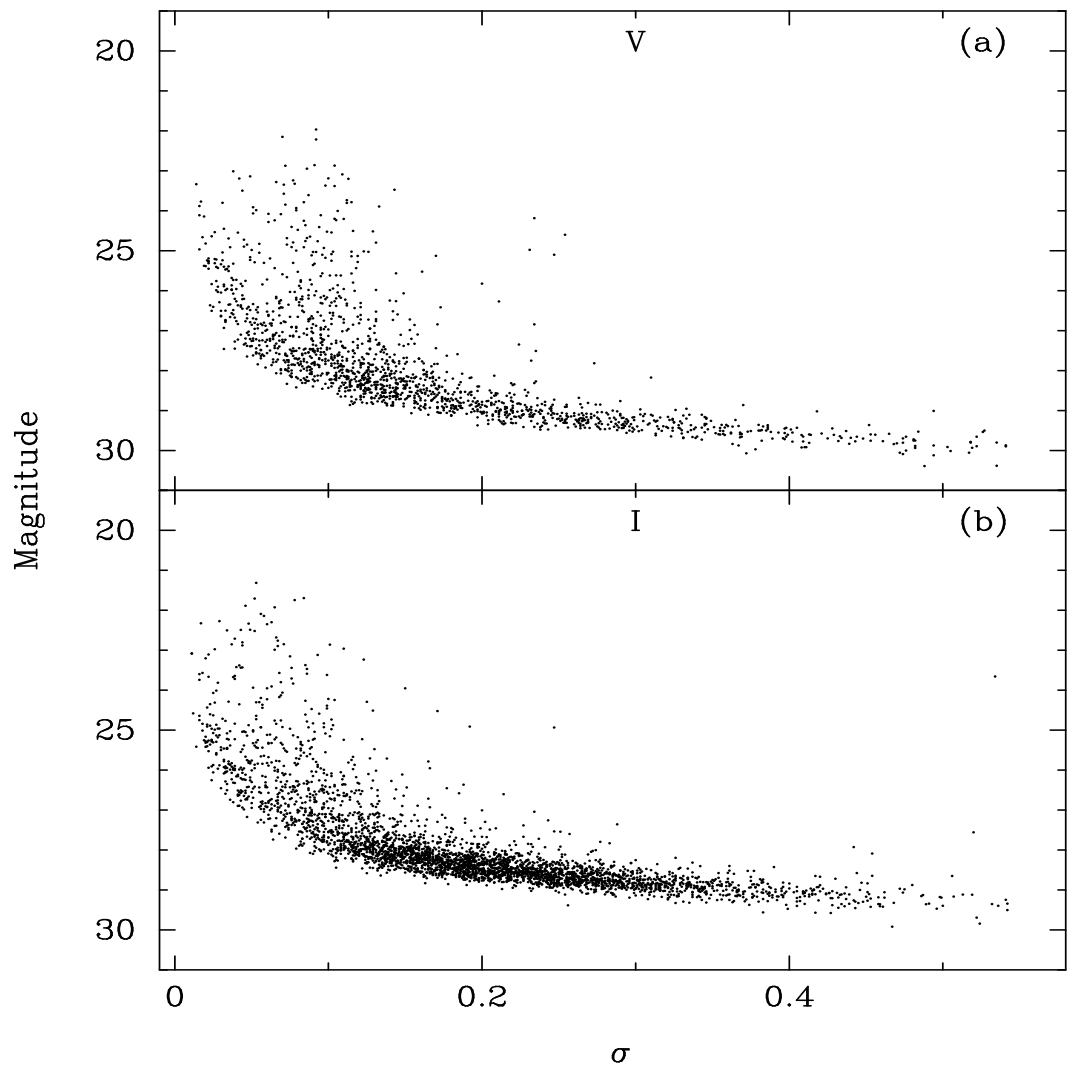}{1pt}{-90.}{400.}{400.}{0.}{0.}
\caption{Photometric error $\sigma$ as a function of apparent $V$ (a) and
$I$ (b) magnitude. The $V$ photometry is
$\sim$ 0.5 mag deeper than the $I$ photometry, going down to a limiting
magnitude of $\sim$ 30 mag.
\label{Fig3}}
\end{figure*}

To minimize the number of false detections, we have adopted the minimum 
detection level to be 3$\sigma$ above the sky noise and a sharpness in the 
--1.0 -- +1.0 range.
Figure \ref{Fig3} shows the distribution of photometric errors as a 
function of $V$ and $I$ magnitudes as determined by DAOPHOT.
It is seen that errors are about 0.2 mag at $V$ = 29 mag and $I$ = 28.5 mag. 
They increase to about 0.4 mag at $V$ = 30 mag and 
$I$ = 29.5 mag. Note that at bright magnitudes ($V,I$ $\la$ 27 mag), the error
distribution is very broad at a given brightness because of the effects of
crowding, ionized gas emission and a high background level of the unresolved
stellar component in the brightest regions of the main body.

The total numbers of recovered stars are respectively 1599, 3555 and 671 in 
the $V$ image, the $I$ image and in both of these images at the same time. 
We have adopted a matching radius of 1 pixel, although changing that 
radius within the range 0.5 -- 2 pixels does not change appreciably 
the number of stars in both $V$ and $I$ frames. 
That more than half the stars are 
not matched is likely caused by incompleteness effects and
an increasing number of false
detections at faint magnitudes. Indeed, the number of the stars with
$V$ = 26 -- 27 mag recovered in both $V$ and $I$ images is
$\sim$ 74\% that recovered in the $V$ image alone, while the number of
the stars with $V$ = 29 -- 30 mag recovered in both $V$ and $I$ images
is only $\sim$ 5\% that recovered in the $V$ image alone.

The transformation of instrumental magnitudes to the Johnson-Cousins $UBVRI$ 
photometric system as defined by \citet{L92} was performed according to the 
prescriptions of \citet{H95}. The magnitudes and colors of
point sources were corrected for Galactic interstellar extinction adopting 
$A_V$ = 0.106 mag \citep{S98}.

\begin{figure*}
\epsscale{2.0}
\plotone{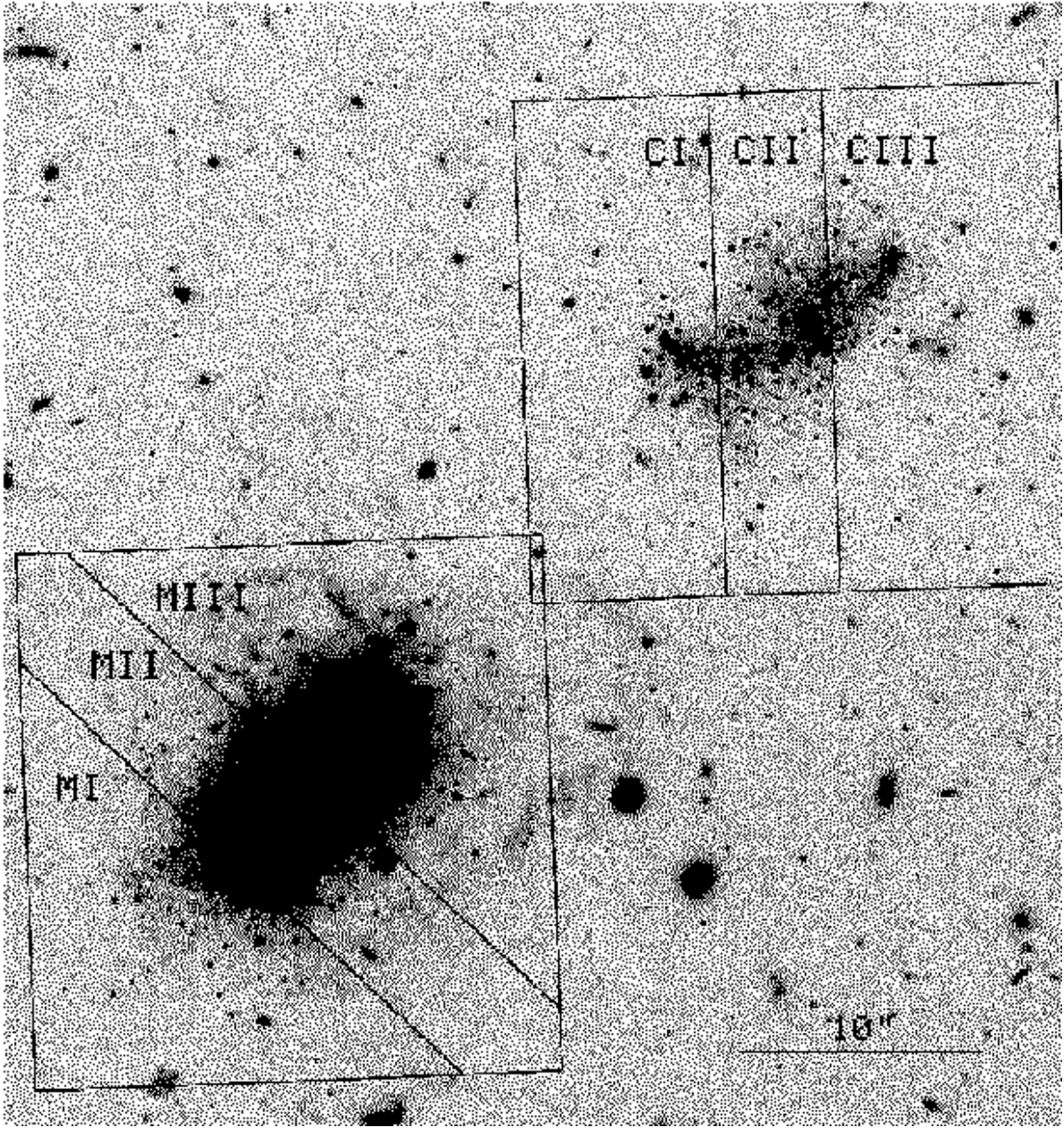}
\caption{$I$ image of I~Zw~18 showing the boundaries and the labeling
of different regions in the main body  (MI, MII, MIII) and in the C component 
(CI, CII, CIII) used for CMD analysis of the stellar populations.
North is up and East is to the left. The scale is shown by a horizontal bar.
\label{Fig4}}
\end{figure*}

We have carried out a completeness analysis using the DAOPHOT routine 
ADDSTAR. Since the crowding level is more important in the main body than
in the C component,
the analysis was done for each component separately, in the regions
determined by the rectangles in Figure \ref{Fig4}.
For each frame we have added artificial stars amounting to $\sim$
5\% the number of real stars detected inside each rectangle.
We then performed a new photometric reduction 
using the same procedure as the one applied to the original frame, and checked 
how many added stars were recovered. This operation was repeated 10
times for each frame and each magnitude bin and the results were averaged. 
The completeness factor in each magnitude bin 
defined as the percentage of recovered artificial stars is shown 
in Table \ref{Tab1}.
It can be seen that the completeness is better for the C component
because of its lesser crowding: 
it is about 
69\% in $V$ and 44\% in $I$ in the 28 -- 29 mag range,
but drops to $\sim$ 17\% and $\sim$ 10\% in the 29 -- 30 mag 
range. For the main body, the completeness is about 
55\% in $V$ and 39\% in $I$ in the 28 -- 29 mag range,
but drops to $\sim$ 10\% and $\sim$ 5\% in the 29 -- 30 mag 
range.

\section{The lack of a RGB stellar population}

\subsection{The color-magnitude diagram of I Zw 18}

\begin{figure*}
\epsscale{1.0}
\plotfiddle{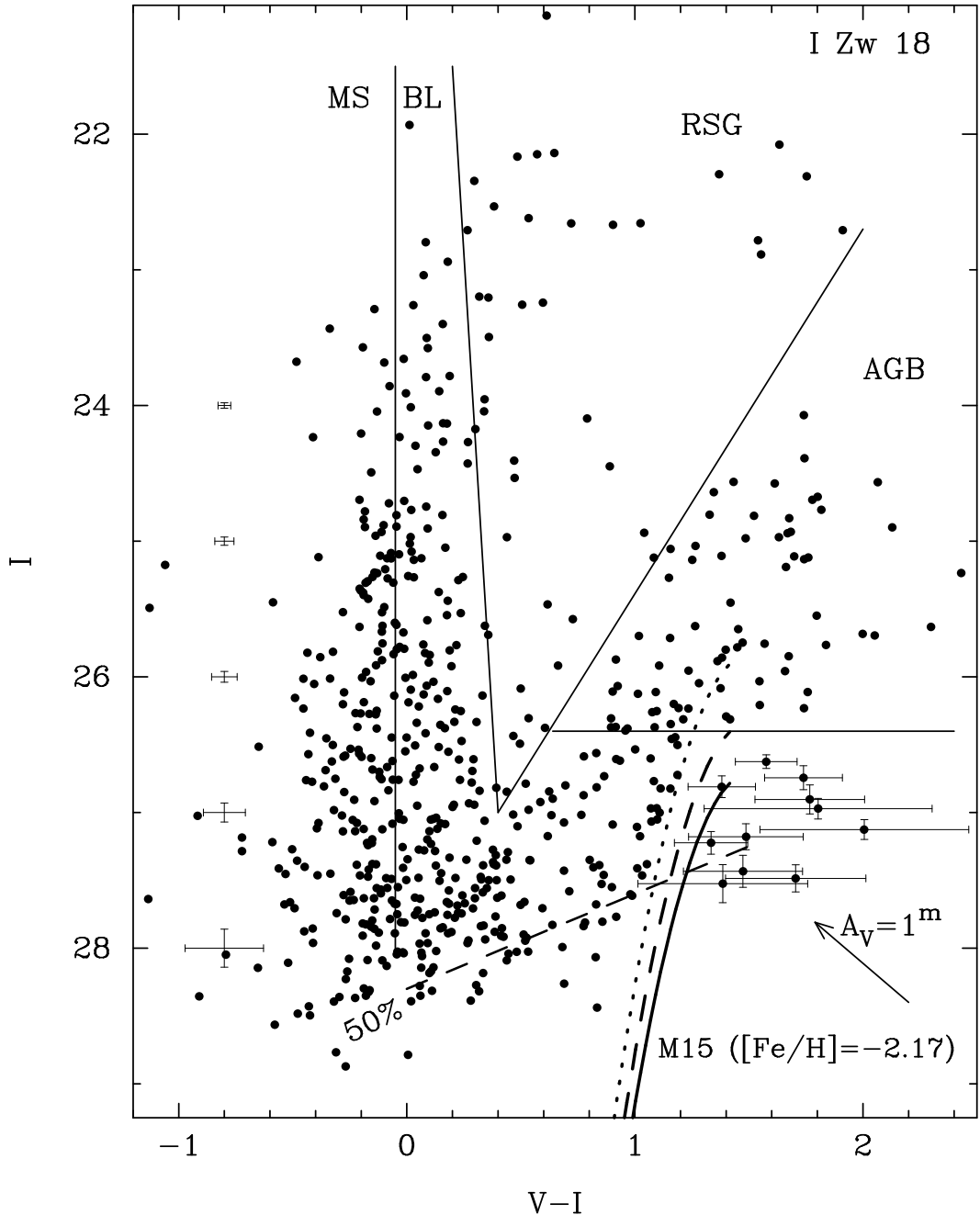}{1pt}{-90.}{500.}{400.}{0.}{0.}
\caption{Combined CMD of both the main body and the C component of
I~Zw~18 overplotted by straight lines showing the locations of the different
stellar types:
main-sequence (MS), blue loop (BL), red supergiant (RSG)
and asymptotic giant branch (AGB) stars. The dashed straight line shows 
the 50\%
completeness limit of stars in both the $V$ and $I$ images.
The thick solid, dashed and dotted 
curved lines are respectively the isochrones of the 
globular cluster M15 with metallicity [Fe/H] = --2.17  from \citet{DA90},
for three distance
moduli $m-M$ = 30.88 mag, 30.5 mag and 30.0 mag, corresponding to distances 
15 Mpc, 12.6 Mpc and 10 Mpc. Mean error bars as a function of $I$ magnitude are
given on the left for MS stars. Error bars for faint ($I$ $>$ 26.6 mag) and
very red ($V-I$ $>$ 1.3 mag) individual sources are also indicated. Also shown
is the extinction vector for $A_V$ = 1 mag.
\label{Fig5}}
\end{figure*}

Figure \ref{Fig5} displays the combined $I$ vs. $V-I$ CMD for 
the main body and the C component. The 50\% detection limit in both $V$ and $I$
filters is shown by a dashed straight line. Our CMD is more than two 
magnitudes deeper than the CMDs obtained earlier for I~Zw~18 
\citep{HT95,D96b,A99,O00}. A well
populated MS can be seen at $V-I$ $\sim$ 0 mag, which begins
with the brightest stars ($I$ $\sim$ 23.5 mag) and goes 
all the way down to 28.5 mag. 
Many bright blue loop (BL) and RSG stars
with $V-I$ $\ga$ 0 mag are present in the upper part of the diagram. 
Two clumps of AGB stars with $I$ $\sim$ 25 mag and $\sim$ 26 mag, and with
$V-I$ redder than 1.0 mag are detected. 
There is, however, a conspicuous lack of a RGB stellar population, despite the
fact that our data, depending on the adopted distance to I~Zw~18, 
goes $\sim$ 1 -- 2 mag fainter than the TRGB. 
Only a few faint red
($V-I$ $>$ 1.3 mag) sources close to the detectability limit  
($I$ $>$ 26.6 mag) are present in the combined CMD. They may be
false detections because of their faintness in $V$ ($\ga$ 28.5 mag).
They may also be real stars with erroneous colors because of the large
photometric uncertainties (see their error bars in Figure \ref{Fig5})
and/or with colors reddened by dust. Their possible nature will be discussed
in Sect. 5.2.3.

We have indicated the expected location of the RGB in the
CMD by superposing on the data points the
isochrone for the very metal-deficient and old globular cluster M15. Its
metallicity [Fe/H] is --2.17 \citep{DA90}, lower than the metallicity of the
ionized gas in I~Zw~18. We have used M15 because it has the lowest
metallicity in the \citet{DA90}' globular cluster sample, and because 
its isochrone fits well the RGB of the very metal-deficient BCD UGC 4483 
($Z_\odot$/23) \citep{IT02}.
Isochrones of globular clusters with higher metallicities are too red
as compared to the data. The dotted, dashed and
solid lines show respectively, from left to right, the location of the M15 
isochrone for three distances of I~Zw~18:
10 Mpc, 12.6 Mpc and 15 Mpc, corresponding to distance
moduli $m-M$ = 30 mag, 30.5 mag and 30.88 mag. 
The upper extremities of each curve are the 
locations of the TRGB at these respective distances. 
It is clear that there is a dearth of sources in the expected region of RGB
stars. This absence is \underline{not} due to a sensitivity effect since it
manifests itself at $\sim$ 1 -- 2 mag above the 50\% completeness line.
Only stars detected at levels greater than 3$\sigma$ above the sky noise in both $V$
and $I$ are shown in Fig. \ref{Fig5}. We have also produced a CMD for I Zw 18 where
the threshould level is only 1$\sigma$ above the sky noise. Examination of that CMD
still reveals a lack of RGB stars in I Zw 18, despite the fact that the CMD goes 
down now 1 mag deeper (limiting $V$ and $I$ magnitudes of above $\sim$ 30 mag).
The CMD of I~Zw~18 is thus unique among all CMDs
of galaxies with resolved stellar populations studied so far
with the {\sl HST}. In those galaxies, 
the RGB stellar population is invariably present and is many times more 
populated than the AGB stellar population 
\citep[see e.g. Figure 14 of][ where the CMDs of 11 dwarf irregular and 
BCD galaxies are presented]{IT02}.

\subsection{Comparison with the resampled color-magnitude diagram of UGC 4483}
 
To quantify our conclusions about the absence of a RGB stellar
population in I Zw 18, we compare the CMD of I Zw 18 with that of a 
nearby galaxy with a well-populated RGB, moved from its actual 
distance to the one of I Zw 18, so that  
the data for the comparison galaxy have a resolution and a sensitivity 
comparable to those of the I Zw 18 observations. 
The cometary blue compact dwarf galaxy UGC 4483 
located at the distance of 3.4$\pm$0.2 Mpc is an ideal galaxy for 
such a comparison. 
With a metallicity of the ionized gas equal to 1/23 that of the Sun, 
it is one of the BCDs with a heavy element content closest to that of I~Zw~18.
It has been resolved into 
individual stars with {\sl HST}/WFPC2 observations by \citet{IT02}.
The RGB stellar population in UGC 4483 with $(V-I)$(TRGB) 
$\sim$ 1.3 mag is one of the 
bluest known among BCDs. Its CMD has been analyzed by \citet{IT02} and is 
given in Fig. \ref{Fig9}c. We discuss in the following how to 
perform the resampling of the data for UGC 4483 when moving it to 
the larger distance of I Zw 18. We need to  
take into account the fading of stars, the increasing errors
of the stellar apparent magnitudes and the increasing crowding effects.

First, we consider how the fading of a star affects  
the probability of its detection.
Assuming that UGC 4483 is at the distance of I Zw 18 and that it 
is observed with the ACS with 
the same exposure times in $V$ and $I$ as for I Zw 18, then 
the detection probability of a star in the CMD of Fig. \ref{Fig9}c
with magnitudes $V$ and $I$ is 
\begin{equation} 
p_m(V+\Delta m,I+\Delta m) = 
p_{\rm ACS}(V+\Delta m)\times p_{\rm ACS}(I+\Delta m),
\end{equation}
where $p_{\rm ACS}$($V+\Delta m$) and $p_{\rm ACS}$($I+\Delta m$) denote the  
incompleteness factors of the ACS observations (Table \ref{Tab1}), 
and $V+\Delta m$ and $I+\Delta m$ are the magnitudes of a star when moved to 
the distance of I Zw 18. Although 
the incompleteness factors 
are different for the main body and the C component, we 
have used only the incompleteness factors
for the latter in our modeling. 
This is because the contamination by ionized
gas emission in UGC 4483 is more
similar to that in the C component than in the main body of I~Zw~18,
where it is considerably higher.

\begin{figure*}
\epsscale{1.5}
\plotone{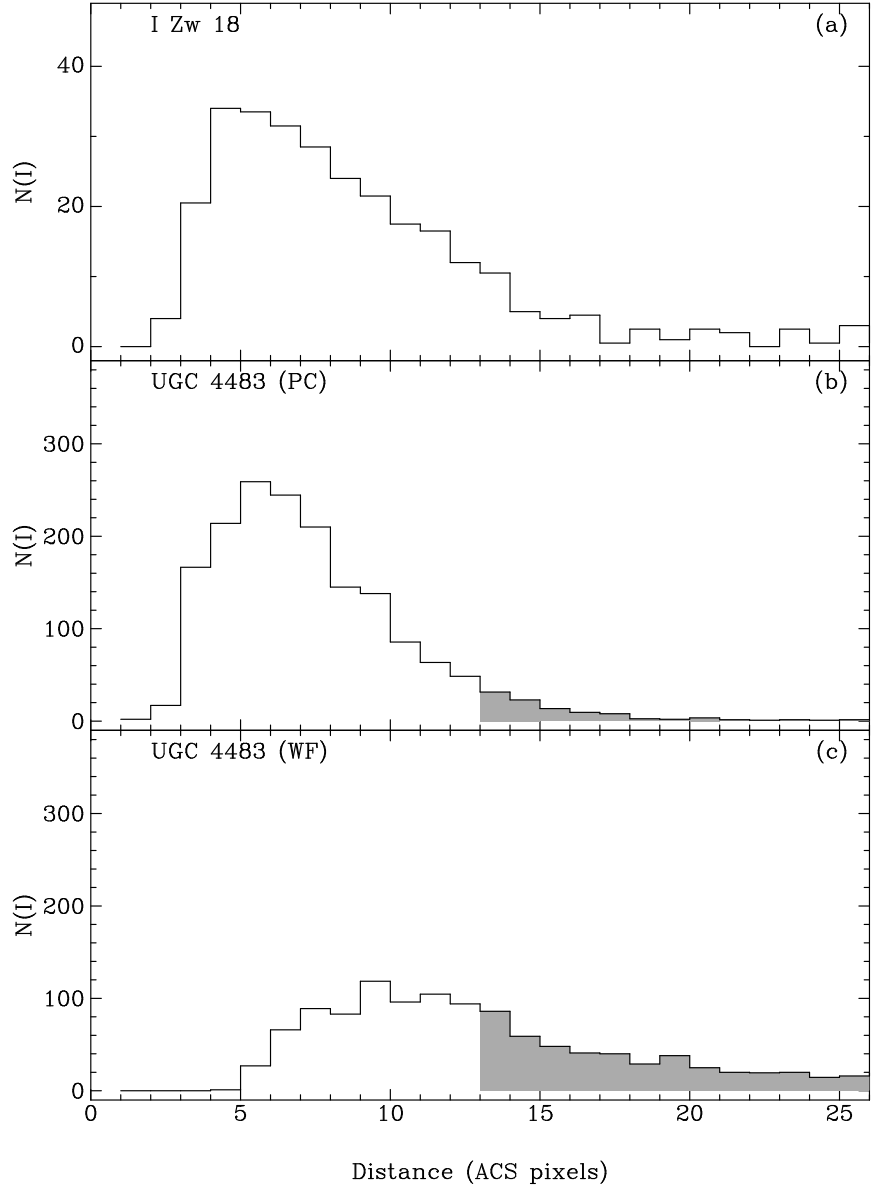}
\caption{Distributions of separations between members of pairs of stars
recovered in both $V$ and $I$ bands in (a) the ACS image of I Zw 18, 
(b) the WFPC2/PC image of the cometary blue compact dwarf galaxy
UGC 4483 ($Z$ = $Z_\odot$/23), and 
(c) the WFPC2/(WF2+WF3+WF4) image of UGC 4483. Data for
UGC 4483 are from \citet{IT02}. The shaded regions in (b) and (c) 
correspond to pairs of stars that are resolved 
in UGC 4483 at the distance of 15 Mpc.
\label{Fig6}}
\end{figure*}

Next we consider the crowding effect: not all stars
that are resolved in UGC 4483 at its distance of 3.4 Mpc 
will be resolved at the distance of I Zw 18. 
We assume that this effect is independent of apparent
magnitude. In Fig. \ref{Fig6} we show the distribution of distances 
 to the nearest companion of each star 
in I Zw 18 (a) and in UGC 4483 (b and c for the parts of the 
galaxy imaged respectively by the PC and by the 
WF2+WF3+WF4 frames, hereafter WF frames). All distances are expressed in 
ACS pixels. It is seen
from Fig. \ref{Fig6}a and \ref{Fig6}b that both stars in a pair 
are resolved by the ACS and the PC only
if their separation is more than 3 ACS pixels. The lower separation 
limits of the ACS and the PC are similar because their resolutions are
nearly the same. On the other hand, in the WFPC2/WF frames, because of 
their twice as worse angular resolution, both stars in a 
pair are recovered only if their separation is more than $\sim$ 
6 ACS pixels.
 Because the linear resolution at a 
distance of 15 Mpc as compared to one of 3.4 Mpc is 15/3.4 = 4.4 times worse,
of the detected stars in UGC 4483, 
only the ones in pairs with separations $\ga$ 3$\times$4.4 $\approx$ 13 ACS 
 will be resolved at the distance of 15 Mpc. The parts of the distributions
occupied by the resolved 
pairs of stars are shown by shaded regions in Fig. \ref{Fig6}b 
and \ref{Fig6}c. They 
correspond to a fraction of stars of $\sim$ 0.06 for the PC frame 
and of $\sim$ 0.51 for the WF frames. 
The remaining pairs with smaller separations
are not resolved. We assume that for those, 
only one of the two stars in a pair is recovered. This  
corresponds to a fraction of stars in unresolved pairs of 0.94/2 = 0.47 for
the PC frame and of 0.49/2 = 0.25 for the WF frames. 

However, these
fractions can be even smaller if there are stars close to the 
unresolved pairs. To estimate this effect, we consider
the distribution of separations between the unresolved pairs of stars and the next
nearest star. We find that only $\approx$ 60\% of unresolved pairs have a 
nearest star with a separation $\ga$ 13 ACS pixels. 
We calculate the fraction
of the stars $p_c$ which is recovered at the distance of 15 Mpc to be: 
0.06 + 0.94/2$\times$0.6 $\approx$ 0.34 for the PC frame,
and 0.51 + 0.49/2$\times$0.6 $\approx$ 0.66 for the WF frames.
In similar manner, we derive $p_c$ = 0.39 (PC) and 0.76 (WF) 
for the distance
of 12.6 Mpc and 0.49 (PC) and 0.87 (WF) for the distance of 10 Mpc.
Finally, the total probability to recover a detected star in UGC 4483 at the 
distance of I Zw 18 is $p$ = $p_m$$\times$$p_c$.

\begin{figure*}
\plotfiddle{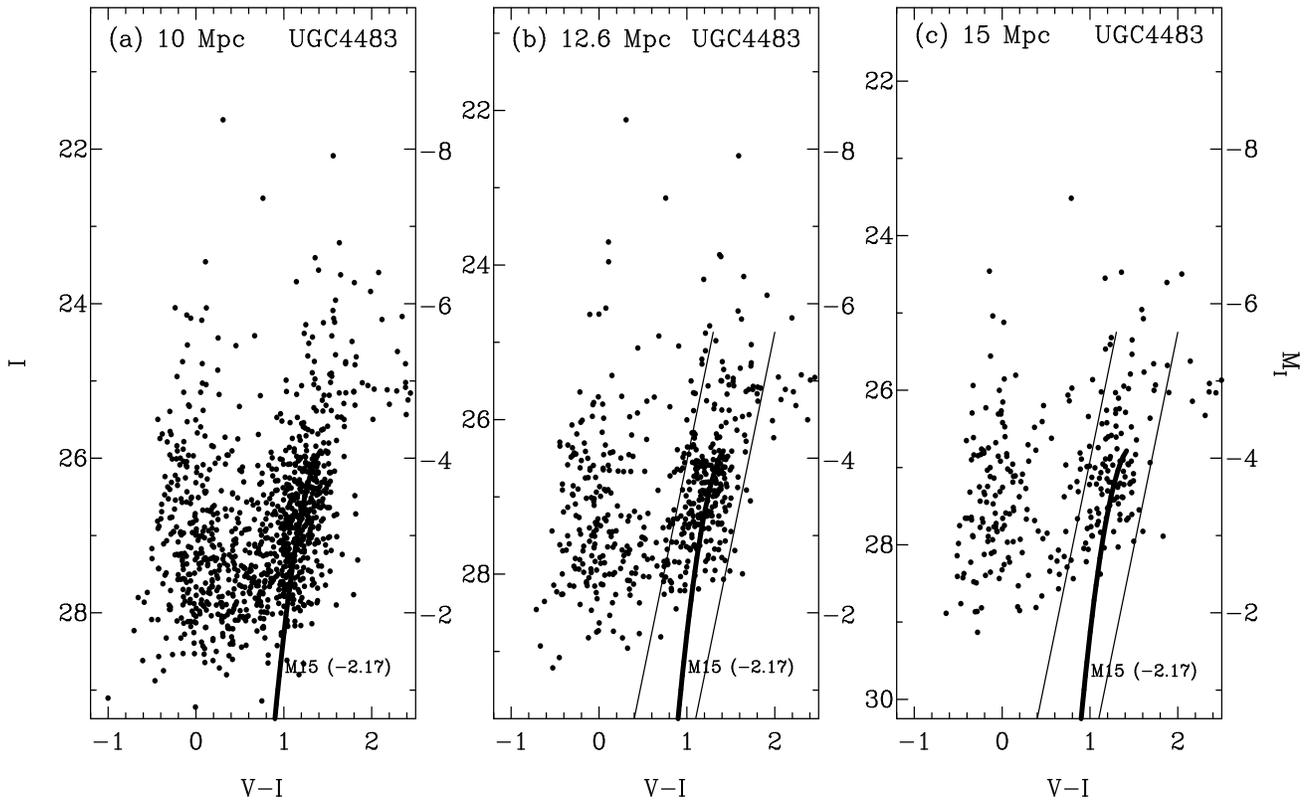}{1pt}{-90.}{300.}{490.}{-10.}{0.}
\caption{Simulated CMDs of UGC 4483 ($D$ = 3.4 Mpc) as they would be seen if 
the galaxy is moved to distances of: 
(a) 10 Mpc (b) 12.6 Mpc and (c) 15 Mpc. The distance-dependent 
effects of incompleteness 
due to star fading and crowding have been taken into account. The left
and right ordinates in each panel are apparent and absolute $I$ magnitudes
respectively. In each panel, the thick line shows the isochrone of 
the globular cluster M15 ([Fe/H] = --2.17), adjusted for the respective 
distances. The two parallel lines in (b) and (c) 
show the CMD region used to construct the 
histograms of number of stars vs magnitude in Fig. \ref{Fig8}.
\label{Fig7}}
\end{figure*}

We show in Fig. \ref{Fig7} representative simulated CMDs of UGC 4483 for 
three possible distances of I Zw 18, 10 Mpc, 12.6 Mpc and 15 Mpc. 
They are generated from the observed CMD 
shown in Fig. \ref{Fig9}c according to the detection probability $p$ 
for each star. 
In addition, we have also recalculated the magnitude of each star 
by taking into account the errors in the magnitude measurements.
For a given stellar magnitude, we have adopted a 1$\sigma$ deviation 
from a fit to the data in Fig. \ref{Fig3} and distributed the errors 
in a gaussian fashion. 
The apparent magnitude scale for each panel 
(left ordinate) has been adjusted so that the absolute 
magnitude scale (right ordinate) is the same in all panels. 
In all three panels, we also show the isochrone for
the globular cluster M15 by a thick line. The two
parallel lines in panels (b) and (c) delineate the location of RGB
stars. We conclude from Fig. \ref{Fig7} that the RGB of UGC 4483
is clearly visible at all three distances. Even if the RGB is sparser at the 
distance of 15 Mpc, it is still clearly there.
In all likelihood, the absence of a RGB in I Zw 18 is not the 
result of a selection effect. 

\begin{figure*}
\epsscale{1.5}
\plotone{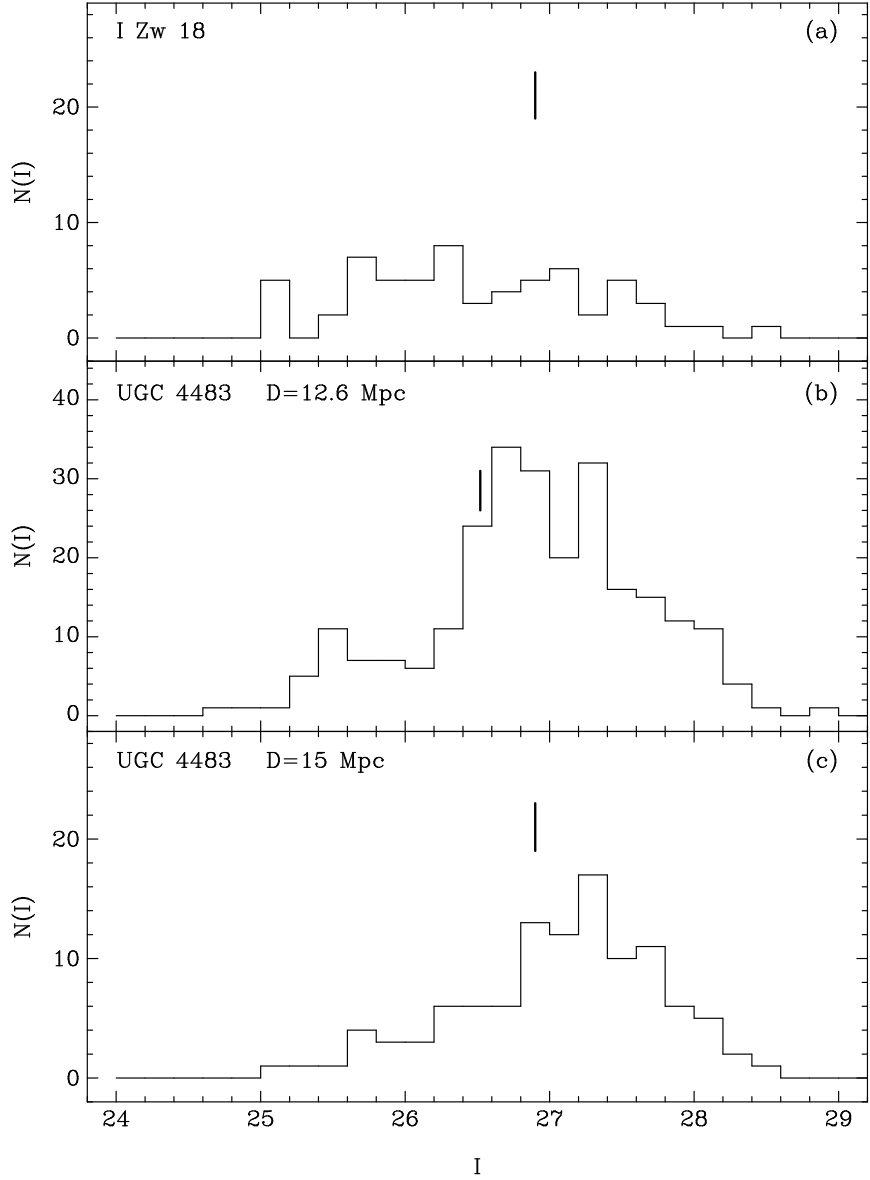}
\caption{ Distributions of numbers of stars vs apparent $I$ magnitude
for the CMD region delineated by the two parallel lines in Fig.
\ref{Fig7} for I Zw 18 (a) and for UGC 4483 at distances of 
12.6 Mpc (b) and 15 Mpc (c), respectively. 
The short vertical lines show the location
of TRGB at the distance of 15 Mpc (a,c) and of 12.6 Mpc (b). 
Note the sharp increase
of the numbers of stars fainter than $I$(TRGB) in UGC 4483 (b and c) due to
RGB stars. No such increase is seen in I Zw 18 (a).
\label{Fig8}}
\end{figure*}

We can quantify the above statement by comparing 
the observed distribution of the numbers of stars between the 
two parallel lines (Fig. \ref{Fig7}b,c and Fig. \ref{Fig9}a,b) as 
a function of $I$ magnitude for I Zw 18 and for UGC 4483 at distances
of 12.6 and 15 Mpc. The distributions 
are shown in Fig. \ref{Fig8}. We mark by short vertical lines the locations
of the TRGB at the corresponding distances. It seen that 
in both galaxies the numbers of AGB stars brighter than the TRGB
are comparable. However, the numbers of stars fainter than the 
TRGB are much larger
in the recalculated CMDs of UGC 4483 than in the CMD of I Zw 18.
A Kolmogorov- Smirnov test shows that, in the 26 -- 28.5
mag magnitude range, 
the observed distribution of stars in I Zw 18 is 
different at the $>$ 99\% and $\sim$ 96\% levels from those in UGC 4483, for 
distances of 12.6 Mpc and 15 Mpc respectively.
For a distance of 10 Mpc, the difference in the distributions of stars is 
at the 99.9\% level.

We have chosen a broad region between the two parallel lines
 for the construction of the histograms shown in Fig. \ref{Fig8}. 
The differences between the histograms of UGC 4483 and that of 
I Zw 18 would be considerably larger 
if we make that region narrower, and in particular if we 
exclude from consideration the stars in I Zw 18 that are concentrated at
the blue edge of the region 0.2 -- 0.3 mag blueward of the M15 isochrone.
There are reasons to think that these blue stars may not be RGB stars. Since 
the metallicity of the stars in I Zw 18
is unlikely to be much lower than the metallicity [Fe/H] = --2.17 
of M15, their blue colors are likely due to an age effect. Hence these stars 
are likely considerably younger than the $\sim$ 10 Gyr old RGB stars in M15. 
Furthermore, the distributions of stars in this region are very different
 in I Zw 18 and in UGC 4483. The RGB stars in the 
CMDs of UGC 4483 are distributed more uniformly and not in a narrow blue edge 
as seen in I Zw 18.

\section{The distance of I~Zw~18}

There has been some debate about the exact distance of I~Zw~18. Values that
have been proposed range from $\sim$ 10 Mpc \citep{HT95,D96b,A99} assuming
the observed heliocentric radial velocity of the galaxy of 
$\sim$ 750 km s$^{-1}$ is a pure Hubble flow velocity and a Hubble constant
of 75 km s$^{-1}$ Mpc$^{-1}$, to $\sim$ 12.6 Mpc \citep{O00} based on a
Virgocentric flow model, to $\sim$ 15 Mpc \citep{I01,IT02} based on 
photometric and spectroscopic properties of I~Zw~18.

With the absence of a RGB stellar population, we cannot use the 
TRGB as a distance indicator.
Other arguments need to be invoked. The redshift distance of I~Zw~18 
without correction for Virgocentric infall is 10 Mpc. If this distance 
is correct, then the $I$ magnitude of the TRGB would be $\sim$ 26 mag,
corresponding to the upper extremity of the dotted line in Figure \ref{Fig5}, 
i.e. it would appear 
brighter than some AGB stars. This seems unlikely as AGB stars are usually
brighter than the TRGB. For example,
the AGB stars in the BCD VII Zw 403 which contains numerous RGB stars are 
$\sim$ 0.5 mag brighter than the TRGB \citep{L98,SL98}. 
In another BCD, UGC 4483, the AGB stars are brighter than the TRGB by 
$\ga$ 0.7 mag \citep{IT02}. Thus, we conclude that a distance of 10 Mpc 
to I~Zw~18 is too small.

There is other evidence that goes in the same sense. \citet{O00}, by 
correcting the radial velocity of I~Zw~18 for Virgocentric infall, 
obtained $D$ = 12.6 Mpc. \citet{I01} and \citet{IT02} suggested that
I~Zw~18 should be as distant as 15 Mpc.
Thus, \citet{I01} found that 
the ionized gas emission seen in the southeastern and central parts of the 
C component can only be produced by stars with ages less than 
$\sim$ 15 Myr or with masses $\ga$ 10 -- 15 $M_\odot$. At a distance of
10 Mpc, the absolute luminosities of the brightest stars in the CMD of 
the C component
would not be large enough for such massive stars \citep[e.g.][]{A99}.
Furthermore, \citet{IT02} have compared the CMDs of 11 galaxies 
(five BCD and irregular 
galaxies outside the Local Group and six Local Group irregular galaxies)
observed with the {\sl HST} and
found that the absolute magnitudes of the brightest stars in 
 the main body of I~Zw~18 would be 
systematically fainter than those in these galaxies if I~Zw~18 is at 10 Mpc,
which is not plausible. 
Increasing the distance of I~Zw~18 to 12.6 Mpc still does not match the 
observations. The youngest bright stars in the CMD of the
C component would still have too low absolute luminosities and too large ages
($\ga$ 25 -- 30 Myr) to account for the ionized gas in the C component. 

\begin{figure*}
\plotfiddle{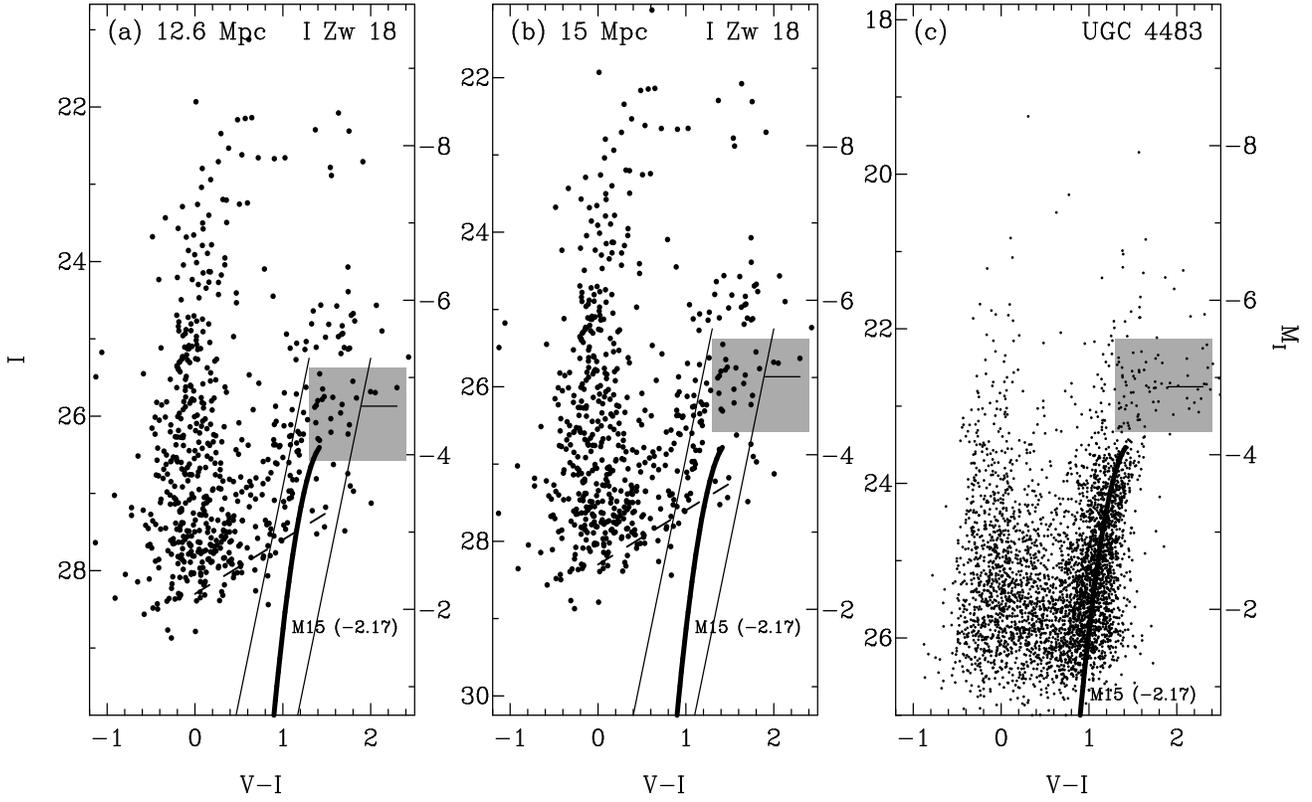}{1pt}{-90.}{300.}{490.}{-10.}{0.}
\caption{Combined CMDs of the main body and the C component for 
two adopted distances of I~Zw~18: (a) 12.6 Mpc and (b)
15 Mpc. (c) Combined CMD of the cometary blue compact dwarf
galaxy UGC 4483 ($Z_\odot$/23) from \citet{IT02}. The left
and right ordinates in each panel are apparent and absolute $I$ magnitudes
respectively. The thick lines in all panels are isochrones of 
the globular cluster M15 ([Fe/H] = --2.17) adjusted for the respective 
distances. The short horizontal lines
indicate the mean absolute magnitude $M_I$ of the AGB stars in the shaded
regions. The dashed lines
represent the 50\% completeness limit of stars in both the $V$ and $I$ images.
The two parallel lines in (a) and (b) show the CMD regions used to construct 
the  
histograms of number of stars vs magnitude in Fig. \ref{Fig8}. 
\label{Fig9}}
\end{figure*}

We estimate the distance to I~Zw~18 by comparing the absolute magnitudes of 
the AGB stars in I~Zw~18 with those in other BCDs.
In Figure \ref{Fig9} we compare the CMDs of I~Zw~18  (Figures \ref{Fig9}a and 
\ref{Fig9}b)
and the BCD UGC 4483 (Figure \ref{Fig9}c). We consider two distances for I~Zw~18, 
12.6 Mpc (Figure \ref{Fig9}a) as proposed by \citet{O00} and 15 Mpc as proposed
by \citet{I01} and \citet{IT02} (Figure \ref{Fig9}b).
 As in Fig. \ref{Fig7}, the 
scales of apparent magnitudes (left ordinate in each
panel) are adjusted so that the absolute magnitude scale (right ordinate in
each panel) is the same in all panels. The thick solid line is 
the isochrone of the globular cluster M15,
the dashed line in Figures \ref{Fig9}a -- \ref{Fig9}b shows the 50\% completeness
limit.  
Two clumps of AGB stars with $I$ $\sim$ 25 mag and
26 mag are seen in the CMD of I Zw 18. For each CMD, we have derived
 the mean magnitudes of the AGB stars
in the shaded regions. In the case of I Zw 18, this corresponds to the fainter 
clump. The short horizontal lines
in Fig. \ref{Fig9} indicate the mean absolute magnitude of
AGB stars thus obtained. 

The absence of RGB stars 
implies that the AGB stars of the fainter clump in I~Zw~18 should be at least 
as bright in absolute magnitude as those in UGC 4483, since they are descendants of more massive stars.
However, at the distance of 12.6 Mpc (Figure \ref{Fig9}a)
the AGB stars in I~Zw~18 are fainter by 0.25 mag than those in UGC 4483 
(Figure \ref{Fig9}c). Their absolute magnitudes are similar 
to those of older AGB stars 
in the BCD VII Zw 403 \citep{L98,SL98}. If we increase the distance of I~Zw~18
to 15 Mpc, then the AGB stars in I~Zw~18 are 0.13 mag brighter, 
a much more satisfactory state of affair
(Figures \ref{Fig9}b and \ref{Fig9}c). 
In fact, the AGB stars in I~Zw~18 and in UGC 4483 have 
the same absolute magnitudes if the distance to I Zw 18 is 14.1 Mpc, 
corresponding
to a distance modulus of 30.75 mag. Poor statistics 
of the AGB stars in the CMD of I Zw 18 and uncertainties in their real 
absolute magnitudes preclude a 
more precise determination of the distance to the BCD. 
All we can say is that it is somewhere between  
12.6 Mpc and 15 Mpc, with the most likely value being   
in the upper range.

\section{Stellar populations}

\subsection{Color-magnitude diagrams}

Our ACS data are considerably deeper than all previous imaging of I~Zw~18, 
obtained with the WFPC2 or NICMOS. For
comparison, the faintest red stars with $I$ $\sim$ 25 mag
and $V-I$ $\ga$ 1 mag
in the CMD of \citet{A99} (their Figure 7b) correspond to the brightest AGB stars
in our CMD (Figure \ref{Fig5}). Likewise, the faintest red stars with
$M_H$ $\sim$ --7.5 mag (at the distance of 12.6 Mpc) 
detected by \citet{O00} correspond to the same
brightest AGB stars in our data. These AGB 
stars are $\sim$ 2.5 mag brighter than the
TRGB absolute magnitude of $\sim$ --5 mag in the $H$ band \citep{F00,V04}.
Furthermore, the ACS observations of the C component have two times better 
spatial resolution as compared to the previous 
WFPC2 observations (0\farcs05 instead 0\farcs1).
Thus, there is no doubt that our deep images allow to study much fainter 
stellar populations and put
more stringent constraints on the evolutionary state of I~Zw~18. 

We have already discussed in section 3 the absence of a RGB stellar 
population in I~Zw~18. We now examine the stellar populations that are 
present.
The CMDs for the main body and C component are shown in Figures \ref{Fig10}a 
and \ref{Fig10}b, respectively. The distance of 15 Mpc to I~Zw~18 is adopted
hereafter. Overplotted on the data are Geneva theoretical isochrones 
(solid lines) of single stellar populations for a heavy element 
mass fraction $Z$ = 0.0004 \citep{LS01}, which corresponds to the metallicity 
of the ionized gas in I~Zw~18. Each isochrone is labeled by the logarithm
of the age
in years.  We choose to use Geneva instead of Padova isochrones \citep{G00}
because the latter are not able to reproduce 
the isochrone of the globular cluster M2 with a similar metallicity, 
[Fe/H] = --1.58 \citep{IT02}, while the agreement is better with Geneva
isochrones.

\subsubsection{The main body}

Several star formation episodes in the main body can be inferred from the CMD
in Figure \ref{Fig10}a and from other observational data. 
The ongoing star formation with age $\sim$ 4 Myr is evidenced
by the ionized gas emission (Figures \ref{Fig2}a -- \ref{Fig2}b) and the
presence of WR stars in the NW component \citep{I97,L97}, located
in two compact clusters \citep{B02}.
The brightest post-MS star in the CMD with an absolute  
magnitude $M_I$ $\sim$ --9.7 mag and an age of $\sim$ 5 -- 6 Myr is also
part of the present burst. 
Numerous supergiants with $M_I$ $\la$ --8 mag indicate that intense
star formation occured $\sim$ 10 -- 15 Myr ago.
There is evidence for two older star formation events in the CMD.
A first star formation episode occurred $\sim$ 200 Myr ago as
indicated by the bright AGB stars with $M_I$ $\sim$ --6 mag.
A second star formation happened 
$\sim$ 300 -- 500 Myr ago, which is responsible for the oldest stars
in the main body, the AGB stars with 
$M_I$ $\sim$ --5 mag. These age estimates are in agreement with the previous
CMD analyses by \citet{HT95}, \citet{D96b} and \citet{A99}, although those
authors discussed only one previous star formation
episode, because their data did not go as deep as ours and the 300 -- 500
Myr old AGB stars were not seen. Most importantly, no old AGB stars 
with ages $\ga$ 1 Gyr are seen, contrary to the assertion of \cite{O00} on the
basis of his {\sl HST}/NICMOS CMD. The presence of such old AGB stars 
would have required numerous RGB stars which are not present
in our CMD. 

\subsubsection{The C component}

\begin{figure*}
\plotfiddle{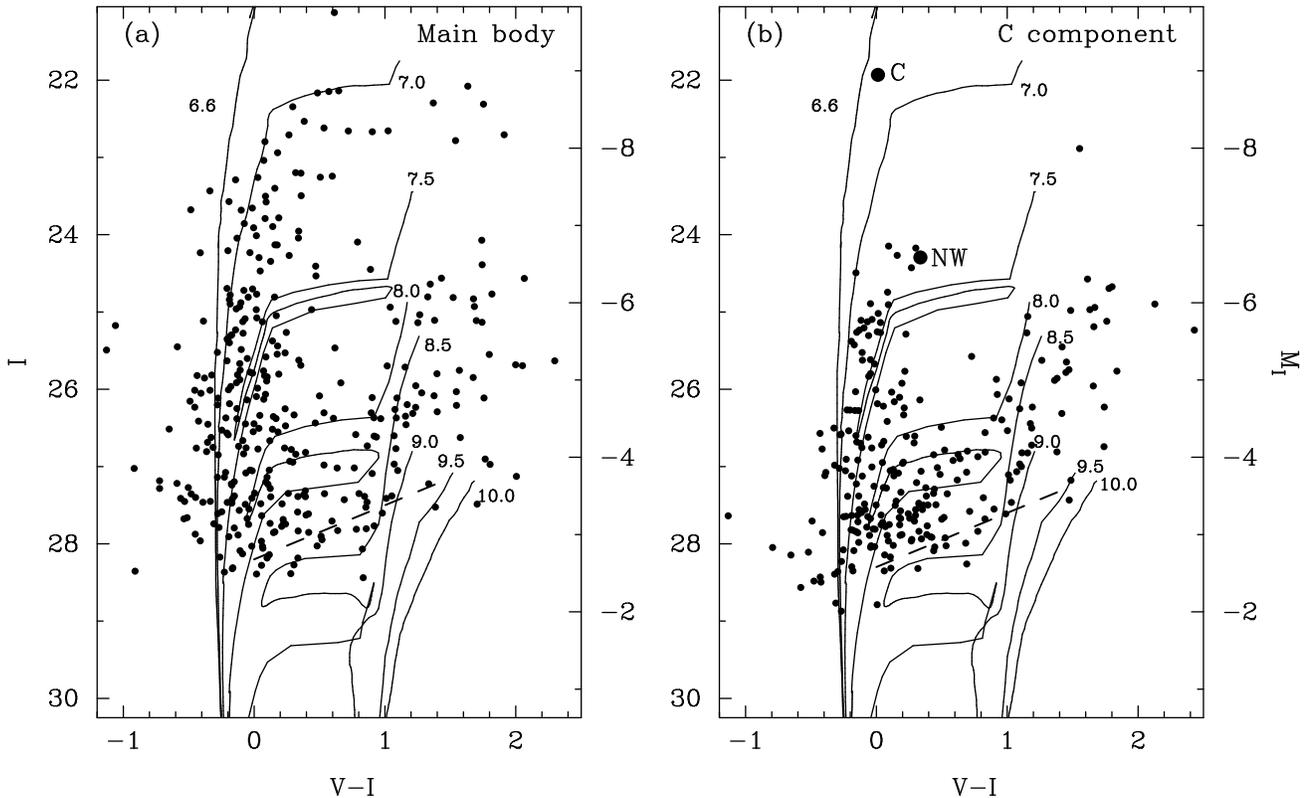}{1pt}{-90.}{300.}{490.}{-10.}{0.}
\caption{$V-I$ vs $I$ CMDs for (a) the main body and (b) the C component
of I~Zw~18, with an adopted distance of 15 Mpc. Left and right ordinates
in each panel are respectively $I$ apparent and absolute magnitudes.
The two stellar clusters in the C component are shown by large filled 
circles and labeled ``C'' and ``NW'' in (b).
Superimposed are Geneva theoretical isochrones for a heavy element mass 
fraction $Z$ = 0.0004 \citep{LS01}. The logarithms of the ages in years 
for each isochrone are shown in each panel. The dashed line represents 
the 50\% completeness limit of stars in both $V$ and $I$ images. 
\label{Fig10}}
\end{figure*}

We turn next to the stellar populations in the C component.
Here, there is no substantial ongoing massive star formation as 
intense ionized gas emission is not seen. This is corroborated by the upper 
part of the CMD, which is nearly devoid of bright stars
(Figure \ref{Fig10}b). The lone bright source and the brightest one in the CMD, 
at $I$ $\sim$ 22 mag
(shown by a large filled circle in Figure \ref{Fig10}b and labeled ``C''
in Figure \ref{Fig2}c and Figure \ref{Fig10}b), 
is in fact not a single star but represents the central 
cluster discussed by \citet{D96b}.
It is partially resolved in our images with a FWHM $\sim$ 0\farcs15, 
corresponding to a linear size of $\sim$ 10 pc. 
There is a second resolved cluster (also shown in Figure \ref{Fig10}b
by a large filled circle and labeled ``NW'' in Figure \ref{Fig2}c
and Figure \ref{Fig10}b) with a FWHM $\sim$ 0\farcs21,
corresponding to a linear size of $\sim$ 15 pc. It was also noted by
\citet{D96b}.

Two star formation episodes can be deduced from the CMD.
The most recent one happened
$\sim$ 15 -- 20 Myr ago while the older one took place
$\sim$ 200 -- 300 Myr ago.
There is an indication that the C component is 
slightly younger
than the main body because its oldest AGB stars are slightly
brighter on average (by $\sim$ 0.2 $I$ mag).
Our age estimates for the C component are also in
agreement with those of \citet{D96b} and \citet{A99} based on
the {\sl HST}/WFPC2 data.

\subsubsection{Individual regions}

\begin{figure*}
\plotfiddle{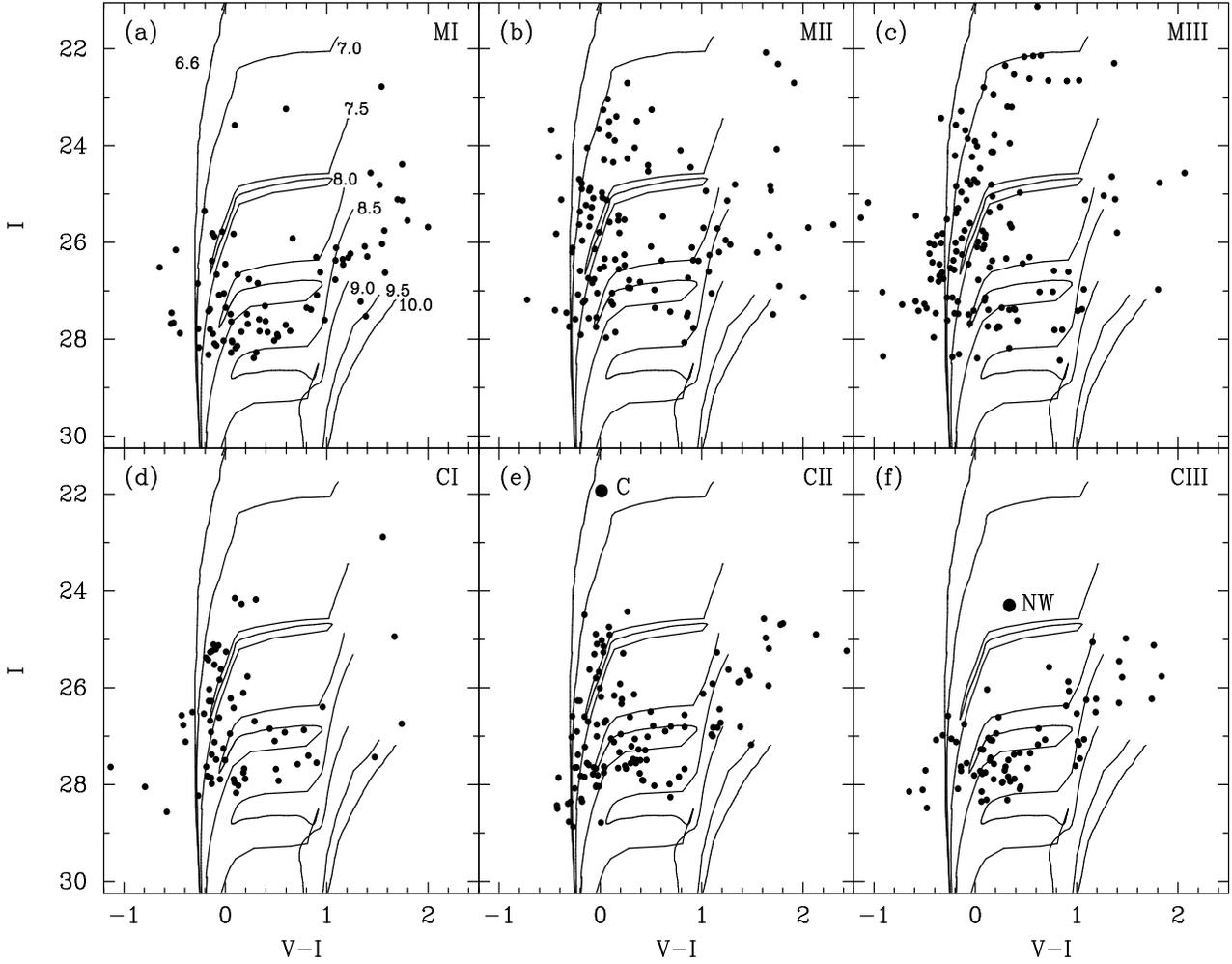}{1pt}{-90.}{375.}{480.}{-10.}{0.}
\caption{$V-I$ vs $I$ CMDs for: (a) - (c) the 3 regions MI, MII, MIII of the 
main body and (d) - (f) the 3 regions CI, CII, CIII of the C component
as delimited in Figure \ref{Fig4}. 
The two stellar clusters in the C component are shown by large filled circles
and labeled ``C'' (e) and ``NW'' (f).
Superimposed are Geneva theoretical 
isochrones for a heavy element mass fraction 
$Z$ = 0.0004 \citep{LS01}. The logarithms of the ages in years 
for each isochrone are shown in (a). 
\label{Fig11}}
\end{figure*}

In Figure \ref{Fig11} we show CMDs for the different regions in the main body 
and in the C component as delimited in Figure \ref{Fig4}. Ongoing and recent
past star formation in the main body is mainly localized in the
MII region associated with the SE component, and in the MIII region 
associated with the NW component.
This is evidenced by the presence of numerous bright
MS (age $\la$ 10 Myr) and BL+RSG (10 Myr $\la$ age $\la$ 100 Myr) 
stars in both CMDs (Figures \ref{Fig11}b -- \ref{Fig11}c). Some AGB stars are also
present in region MII. On the other hand, only a few RSG
stars are seen in the CMD of the southernmost region MI. The population
of AGB stars is more numerous there, indicating older star formation
(age $\ga$ 100 Myr) (Figure \ref{Fig11}a).

In the CMD of the southernmost region CI of the C component (Figure \ref{Fig11}d)
only stellar populations with ages $\la$ 100 Myr are present. 
The age of
the youngest stars in this region is $\sim$ 15 -- 20 Myr as evidenced by
the brightest post-MS stars ($M_I$ $\sim$ --8 mag)
in the CMD  and the presence of weak H$\beta$ and H$\alpha$ emission
lines in its spectrum \citep[e.g.,][]{I01}. In the central region CII 
(Figure \ref{Fig11}e), older stars with 
ages $\sim$ 200 -- 300 Myr are present. 
There is a population of younger stars, with ages $\sim$ 30 Myr as deduced 
from the $M_I$ $\sim$ --6.4 mag of the brightest post-MS
stars. Although they are not resolved, the youngest stars (age $\sim$ 15 Myr) 
are likely present in
the central cluster, as evidenced by the presence of ionized gas emission 
there \citep{D96b,I01}. Region CIII is the oldest (age $\ga$ 300 Myr) 
region in the C component (Figure \ref{Fig11}f).
There is no recent star formation here since no bright MS star is seen.
The AGB stars are fainter than those in region CII, also suggesting
a larger age. Probably, the youngest object in region CIII is the 
NW stellar cluster. However, the absence of ionized gas emission around
this cluster \citep{D96b} and its relatively red $V-I$ $\sim$ 0.35 mag
(Figure \ref{Fig11}f) implies that its age is $\ga$ 100 Myr.

\subsubsection{The age of I~Zw~18}

All previous CMD studies of I~Zw~18 did not go deep enough to allow setting
an upper limit to its age. Our new deep ACS images permit us to do so.
For a distance of I Zw 18 in the range between 12.6 Mpc and 15 Mpc, 
no RGB stars are seen in our CMDs, and the age upper 
limit can be set to 1 -- 2 Gyr. However, the real age for I~Zw~18
is likely smaller as only stars with ages $\la$ 500 Myr are present 
(Figures \ref{Fig10} and \ref{Fig11}). This age upper limit of $\sim$ 500 Myr 
is in excellent agreement with the age estimates obtained by \citet{P02} and 
\citet{H03} by examining the integrated optical and near-infrared colors 
of I~Zw~18.

While we favor a distance of I Zw 18 in the upper range of the 12.6 Mpc -- 15 
Mpc interval, our conclusions will be reenforced 
if the smaller distance of 12.6 Mpc
is adopted for the BCD. In this case, it would be easier to detect faint RGB 
stars since the TRGB's apparent magnitude would be
up to 0.38 mag brighter
(compare Figures \ref{Fig9}a and \ref{Fig9}b with the simulated CMDs for
UGC 4483 in Figures \ref{Fig7}b and \ref{Fig7}c). 
The ages of the oldest stellar
populations will increase at most by a few hundred Myr. However,
we have argued in section 4 that such a small distance would not be in 
agreement with 
photometric and spectroscopic observations of I~Zw~18.

\subsection{Spatial distributions}

The spatial distributions of stars with different ages in I~Zw~18 
can give useful information
on its star formation history. To carry out the study, we divide the
stars in the CMD in Figure \ref{Fig5} into three categories: 
1) MS stars 
with $V-I$ $<$ --0.05 mag, 2) BL stars with 
--0.05 mag $\leq$ $V-I$ $<$ 0.4 mag and
RSG stars with $I$ $<$ 24 mag and $V-I$ $\geq$ 0.4 mag, and 3) AGB stars 
with 26.4 mag $\leq$ $I$ $\leq$ 24 mag and $V-I$ $\ga$ 0.8 mag. 
We have also included in the last category the
progenitors of the AGB stars located in the lower right corner of the CMD with
26.4 mag $\leq$ $I$ $\leq$ 27.5 mag and 0.6 mag $\leq$ $V-I$ $\leq$ 1.2 mag. 
The boundaries of the regions in the CMD of I~Zw~18 where these stars are 
located are shown in Figure \ref{Fig5}.

The spatial distributions of different types of stars are displayed
by in Figures \ref{Fig12}a -- \ref{Fig12}d. The open circles indicate the locations 
of the NW and SE
components in the main body and the central star cluster in the C component.
The spatial distribution of all stars irrespective of type is shown in 
Figure \ref{Fig12}a. It is seen that the vast majority of the stars are 
located in
the main body and in the C component. The few scattered points 
outside I~Zw~18 are probably false detections.

\subsubsection{The main body}

\begin{figure*}
\plotfiddle{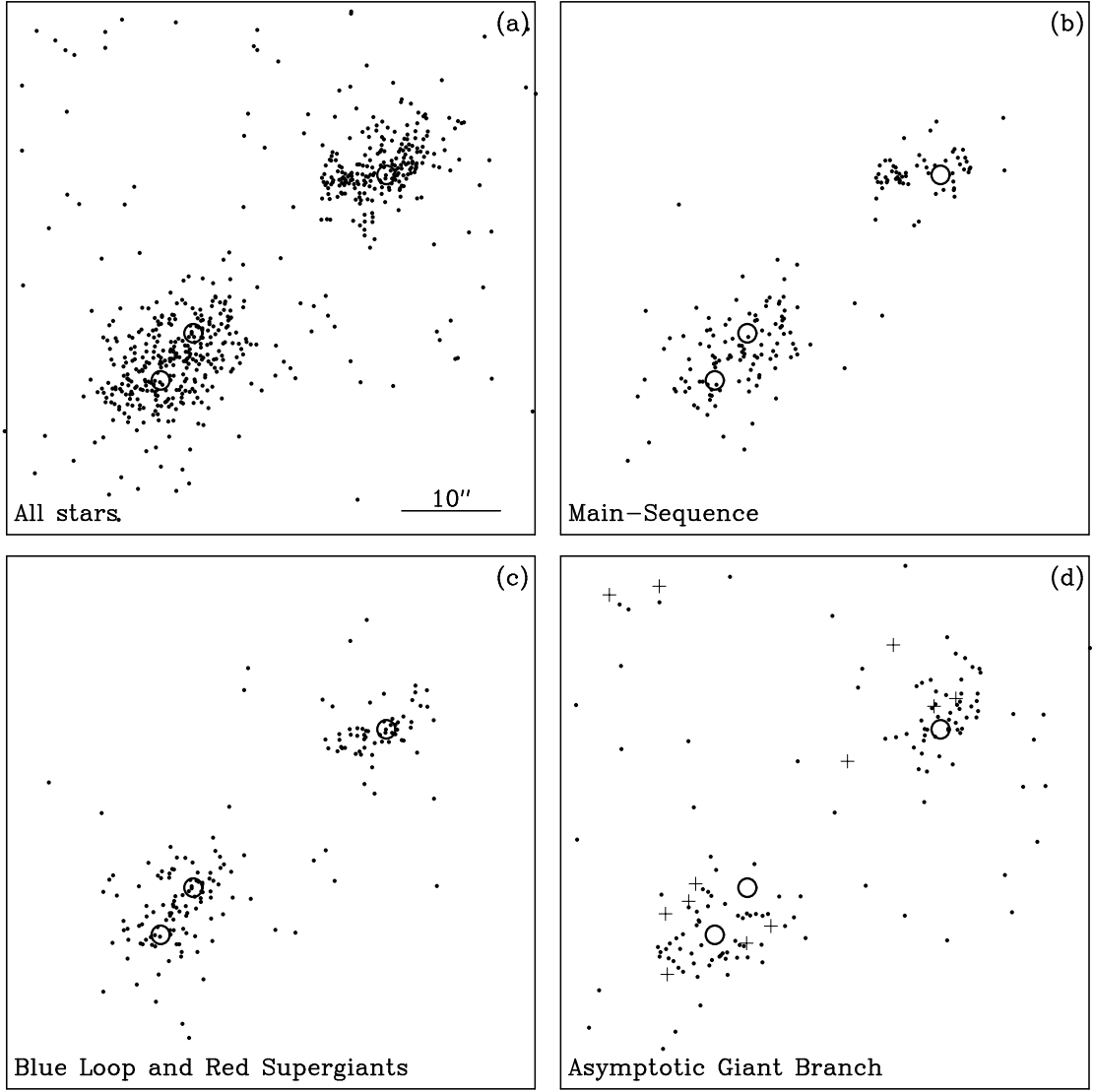}{1pt}{-90.}{420.}{420.}{0.}{0.}
\caption{Spatial distributions of 
(a) stars of all types, (b) MS stars, (c) BL 
and RSG stars and (d) AGB stars in I~Zw~18
(dots). Crosses in (d) show faint red stars with
$I$ $>$ 26.6 mag and $V-I$ $>$ 1.3 mag (see the points with error bars
in Figure \ref{Fig5}).
Open circles show the locations of the NW and SE components in the main body
and of the central cluster in the C component. In the main body, the younger
MS and BL+RSG stars are distributed in larger areas as compared to the older
AGB stars [compare (b) and (c) with (d)]
suggesting that I~Zw~18 is building up from the inside out. 
In the C component, stars of different ages are located in different
regions reflecting the stochastic mode of star formation in BCDs and also
suggesting ongoing formation of I~Zw~18. 
The scale is shown in (a). North is up and East is to the left.
\label{Fig12}}
\end{figure*}

In the main body, the MS and 
BL+RSG stars (Figures \ref{Fig12}b -- \ref{Fig12}c) are distributed 
uniformly, suggesting that ongoing and recent past
star formation took place over the
whole body. On the other hand, the AGB stars are mainly located in 
the southeastern part
(Figure \ref{Fig12}d). This was noted earlier by \citet{A99} and
\citet{IT02} based on WFPC2 images. The
absence of AGB stars in the northwestern region may be caused in part by a 
more severe crowding and a larger
contribution of ionized gas emission in the $V$ image
(Figures \ref{Fig2}a -- \ref{Fig2}b). This may make DAOPHOT miss more of the
AGB stars since they are fainter in $V$ than the MS
and BL+RSG stars. However, we believe that some of the effect is real.
The most striking feature is
that the older AGB stars are not more spread out spatially than the younger 
stars, a fact already noted by \citet{IT02}. Such a distribution is 
drastically different from the situation in other galaxies
where the old AGB and RGB stars are distributed over a considerably larger 
area as compared
to younger stars, due to diffusion and relaxation processes of stellar
ensembles. If anything, the reverse appears to be true here: the
MS and BL+RSG stars are distributed over a
larger area around the main body as compared to the AGB stars. 
This suggests that star formation in the past responsible for the AGB stars
was more concentrated in the main body,
while recent star formation responsible for the MS and BL+RSG stars 
is more spread out. Evidently, I~Zw~18 is a galaxy in the process of forming
from the inside out.

The absence of a halo of AGB stars and of their progenitors around I~Zw~18,
despite an easier detectability at large distances because of less crowding,
is in direct contradiction with the scenario
of \citet{L00a} and \citet{L00b}. These authors have proposed that continuous 
low-mass star formation has proceeded in I~Zw~18 on
cosmological time scales in a large area around the main body.
They predict the existence
of a red extended underlying stellar component which \citet{KO00} claim to
have detected from
their optical and near-infrared surface photometry, but which was not confirmed
by independent photometry by \citet{P02} and \citet{H03}.  
In fact, the spatial distributions in Figure \ref{Fig12} 
suggest that the AGB stars in the main body are relatively young and have had
no time to migrate to large distances \citep[cf.][]{IT02}.

\subsubsection{The C component}

In the C component, the spatial distributions of the 
different types of stars are very
different from those in the main body. The MS stars are seen primarily
in the southeastern part and are nearly absent in the 
northwestern part (Figure \ref{Fig12}b). The BL+RSG stars are
also mainly distributed in the southeastern part, however
their location is offset as compared to the MS stars: the MS stars are mostly
aligned in the East - West direction while the BL+RSG stars are mainly 
distributed in the southeast - northwest direction. On the other hand, AGB
stars are located principally 
in the northwestern part. Some of these stars are also present to the
south of the central cluster. It appears
that the two major episodes of star formation in the C component have
occurred in spatially different regions. This is characteristic of the mode of
star formation in BCDs where the centers of star formation move about in a
stochastic manner. This also implies that, as for the main body, 
the formation of
the C component is still proceeding. It started in the northwestern
part $\sim$ 200 -- 300 Myr ago, lasting for a relatively short period,
and continues now in the southeastern part. That MS stars are not seen
in the northwestern part does not mean that such stars are not
present there. They are simply too faint to be detected because the MS
turn-off for a 200 -- 300 Myr old stellar population is $M_I$ 
$\sim$ --1 mag \citep[e.g.][]{G00} or $I$ $\sim$ 30 mag, below the 
detectability limit. The absence of MS stars also means that
no detectable star formation has occurred during the 
last $\sim$ 200 Myr in the northwestern
part. One possible exception is the NW cluster with the age $\ga$ 
100 Myr. On the other hand, the absence of AGB stars in
the southeastern region, where MS stars reside, suggests that 
this region is younger, with an age
$\la$ 100 Myr. A similar age estimate
was made by \citet{I01} from a spectroscopic study.

\subsubsection{Faint red stars}

Finally, we consider the spatial distribution and nature of the few faintest 
($I$ $>$ 26.6 mag) and reddest ($V-I$ $>$ 1.3 mag) sources 
in the CMD of Figure \ref{Fig5}. They
are shown by crosses in Figure \ref{Fig12}d. A few of these sources are
scattered in the field and are likely false detections. Two other are
in the C component, while the remaining six are in the southeastern
part of the main body. Their $V-I$ colors are redder than those of globular
cluster stars in M15. The location of some of these sources
in the CMD of the main body to the right of the 10 Gyr Geneva theoretical 
isochrone with $Z$ = 0.0004 (Figure \ref{Fig10}a) implies that they are 
older than the Universe, which is absurd. The situation is worse
when Padova isochrones are used.
Thus, we conclude that the red colors of the faint sources are not due to
their large ages, but to some other reason.
Their very red $V-I$ colors may be explained in part by large
photometric uncertainties, which at these faint magnitudes, can reach
$\la$ 0.5 mag (see the error bars in Figure \ref{Fig5}). 
Furthermore, these sources may be reddened by dust
known to be present in the main body, with a maximum extinction $A_V$ 
$\sim$ 0.5 mag in the southeastern part
\citep{C02}. 
However, this value is derived over large areas, and the extinction can be
clumpy and significantly higher on small scales.
It is seen from Figure \ref{Fig5} that correction for 
interstellar reddening with $A_V$ $\ga$ 1 mag will move all red faint sources
in the CMD region where stars with ages $\la$ 500 Myr reside.
In any case, the faintness of these sources and the unknown distribution
of extinction on small scales in the main body and in the C component 
precludes a more reliable explanation of their nature.

\subsection{Surface-brightness and color distributions}

   Another way to study the properties of stellar populations is to consider
their integrated characteristics as given by the surface brightness and color
profiles in different regions of the galaxy. The advantage of this approach 
is that it includes both resolved and unresolved stars. The disadvantage is 
that populations
with different ages contribute to the integrated light and assumptions have to 
be made on the star formation history to derive the age distribution of stars.

\begin{figure*}
\epsscale{1.6}
\plotone{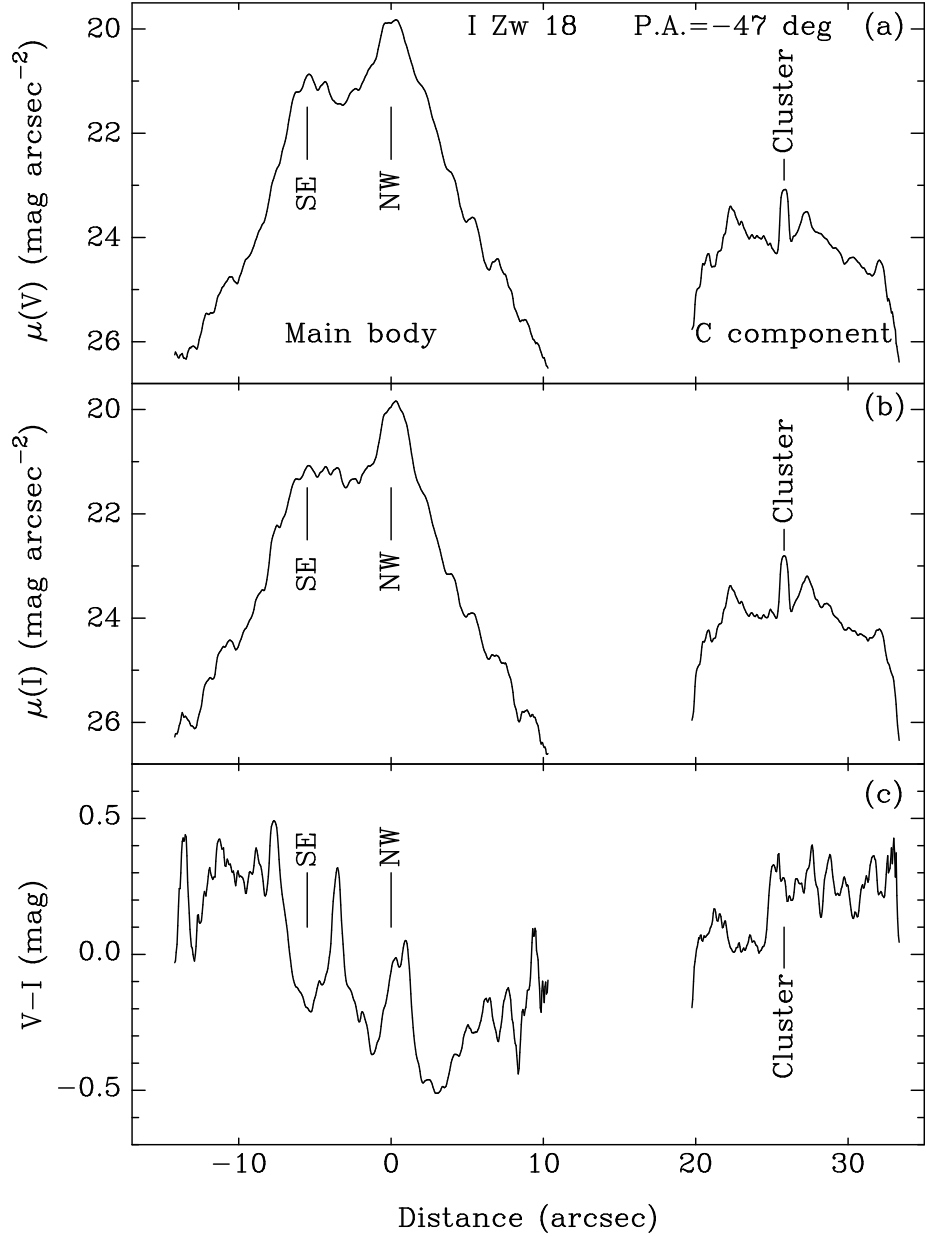}
\caption{$V$, $I$ and $V-I$ distributions in a 7\arcsec-wide strip
with position angle --47\arcdeg\ 
connecting the main body and the C component. The origin is taken to be at the 
location of the NW component in the main body.
\label{Fig13}}
\end{figure*}

   In Figure \ref{Fig13}, we show the $V$ (a) and $I$ (b) surface-brightness
and $V-I$ (c) color distributions averaged over a 7\arcsec\ wide strip along 
the position angle --47\arcdeg\ connecting the main body and the C component. 
The origin is taken to be at the center of the NW component of the main body.
Surface brightnesses and colors have been transformed to the
standard $VI$ photometric system according to the prescriptions
of \citet{H95} and have been corrected for Galactic extinction with 
$A_V$ = 0.106 mag \citep{S98}. The NW and SE components of the main body
and the central cluster in the C component are labeled in each panel.
The bluest color $\sim$ --0.5 mag is in a region $\sim$ 3\arcsec\ northwest
of the NW component. The equivalent width
of the H$\alpha$ emission line in this region exceeds 1000\AA\ \citep{I01} 
and hence the blue
color is due to ionized gas emission. Other regions in the main body
with $V-I$ as blue as $\sim$ --0.3 - --0.4 mag are also strongly contaminated
by ionized gas emission. The only region in the main body free of
ionized gas emission is the southernmost one (region MI) which, 
with $V-I$ $\sim$ 0.3 -- 0.4 mag, is the reddest part
in I~Zw~18. In the C component, the 
contribution of ionized gas emission is negligible \citep{I01} and the bluest
color $V-I$ $\sim$ 0 mag in the southeastern part (region CI) 
really reflects that of a
young stellar population. The reddest  color $V-I$ $\sim$ 0.2 -- 0.3 mag 
is in the northwestern part of the C component. It is slightly bluer than the
color of the reddest region in the main body, suggesting that the C 
component may be slightly younger than the main body. The surface brightness
profiles also show the absence of an extended LSB component,
with a precipitous drop of 6 mag from the NW component to the edge of the
galaxy over a distance $\sim$ 10\arcsec\ or 750 pc.

To model the colors of the reddest regions in I~Zw~18, we use spectral 
energy distributions of single stellar populations with a heavy element mass 
fraction $Z$ = 0.0004, calculated with the PEGASE.2 code of \citet{FR97}.
Because Geneva stellar evolutionary models \citep{LS01} do not include AGB 
stars, the Padova models 
\citep{G00}\footnote{http://pleiadi.pd.astro.it.} are used.
The Salpeter initial mass function (IMF) with a slope $\alpha$ = --2.35, 
and lower and upper stellar mass limits
of 0.1 $M_\odot$ and 120 $M_\odot$ respectively are adopted.
The models predict that the $V-I$ color of an 
instantaneous burst with age between 200 and 500 Myr is 
$\sim$ 0.5 mag, or $\sim$ 0.1 -- 0.2 mag redder
than the observed reddest color. Only for an instantaneous burst with age 
$\la$ 150 Myr, several 100 Myr smaller than the CMD-derived ages, 
are the modeled colors consistent with the observed ones.
We have also considered the case of continuous star formation.
If star formation has occurred at a constant rate 
continuously during the period 
50 -- 160 Myr ago, then the $V-I$ color is $\sim$ 0.35 mag, in agreement 
with the observed value. But the
upper age limit is again significantly lower than the CMD-derived ages.
We can increase the upper age limit for continuous star formation 
to make it consistent with the CMD-derived value of $\sim$ 500 Myr, but
to reproduce $V-I$ $\sim$ 0.35 mag, we must at the same time decrease the
lower age limit to $\sim$ 10 Myr. In this case young massive stars should 
be present in the reddest regions of I~Zw~18, and they are not seen. 

A possible cause for the color discrepancy may be 
that the metallicity of stars in I~Zw~18 is not equal to the metallicity of
the ionized gas as assumed, but lower. We have checked into this possibility
with models that are four times as metal-poor ($Z$ = 0.0001). $V-I$ does become
bluer, but the blueing is not sufficient. For instantaneous bursts with ages in
the range 200 -- 500 Myr, $V-I$ $\sim$ 0.46 mag, still too red as compared
to the observed value. We have also varied the parameters of the IMF
for a single stellar population with a heavy element mass fraction 
$Z$ = 0.0004.
Making the slope flatter or steeper with $\alpha$ in the range --2 - --3 and 
increasing the lower 
mass limit from 0.1 $M_\odot$ to 1 $M_\odot$ does not result in a significantly
bluer $V-I$ color. The $V-I$ colors are in the range 
0.45 -- 0.50 mag for instantaneous bursts with ages in the range 
200 -- 500 Myr. We conclude from the previous considerations that,
despite the uncertainties of the models and the fact that they cannot
reproduce precisely the observed reddest colors in I~Zw~18, they do 
predict young ages which are in the ballpark of CMD-derived ages.
    
\section{Summary}

   We present a photometric study of the resolved stellar populations in
I~Zw~18, the most metal-deficient blue compact dwarf (BCD) galaxy known.
The analysis of the color-magnitude diagram (CMD) of I~Zw~18, based on 
{\sl Hubble Space Telescope}/Advanced Camera for Surveys 
$V$ and $I$ images, the deepest ever obtained
for the BCD,  have led us to the following conclusions:

   1. The CMD of I~Zw~18 is populated by stars with different ages
including the youngest hydrogen core burning main-sequence (MS) stars
(age $\la$ 30 Myr), evolved massive 
stars with helium core burning [blue loop (BL) stars and red supergiants
(RSG)] with ages between 10 Myr and 100 Myr,
and asymptotic giant branch (AGB) 
helium shell burning stars with ages between 100 Myr and 500 Myr. 
However, {\it I~Zw~18 is the
first galaxy with resolved stellar populations where no red giant branch
(RGB) stars are seen}, although our data go 1 -- 2 mag deeper than the tip
of the RGB for a distance of I Zw 18 in the range 12.6 Mpc -- 15 Mpc (see 
conclusion 2). 
The oldest stars located mainly
in the southeastern part of the main body and the northwestern part of the 
C component have an age not exceeding $\sim$ 500 Myr.
Thus {\it I~Zw~18 is a bona fide young galaxy}. 

   2. Since no RGB stars are seen, we cannot use the brightness of the tip of 
the RGB to derive the distance to I~Zw~18. Instead, we compare the brightness 
of the AGB stars in I~Zw~18 with those in the very metal-deficient BCD 
UGC 4483 with a heavy element abundance of $Z_\odot$/23 to derive a distance
in the range 12.6 Mpc -- 15 Mpc, with the most likely value in the 
upper range.

   3. Several star formation episodes in I~Zw~18 can be inferred from
its CMD. However star formation
proceeds differently in the main body and in the C component.
In the main body, three star formation episodes are indicated,
separated by periods of $\sim$ 100 -- 200 Myr.
Examination of the spatial distribution of the stellar populations suggests
that the star formation process is still gradually building up the
main body from the inside out as the young MS and BL+RSG stars
occupy larger areas as compared to the older AGB stars.
In the C component, two star formation episodes are inferred
separated by a period of
$\sim$ 200 Myr. These separate star formation episodes occurred in spatially
different regions, reflecting the stochastic mode of star formation in BCDs. 
The southeastern region of the C component is
a few hundred Myr younger than the northwestern region. 
Both in the main body and in the C component, the spatial
distributions of the stellar populations in I~Zw~18 strongly suggest that the
galaxy is still in the process of forming.

\acknowledgments
T.X.T. has been partially supported by grant HST-GO-08769.01-A.
The research described in this publication was made possible in part by Award
No. UP1-2551-KV-03 of the U.S. Civilian Research \& Development Foundation 
for the Independent States of the Former Soviet Union (CRDF) and a grant
No. M/85-2004 of the Ministry of Education and Science of Ukraine.
We are also grateful for the partial financial support of NSF grant 
AST-02-05785.
Y.I.I. thanks the hospitality of the Astronomy Department of the University of 
Virginia.


%
\begin{deluxetable}{crrcrr}
\tablenum{1}
\tablecolumns{6}
\tablewidth{0pt}
\tablecaption{Photometry completeness\tablenotemark{a}
\label{Tab1}}
\tablehead{ Magnitude & \multicolumn{2}{c}{Main body}&
& \multicolumn{2}{c}{C component} \\
\cline{2-3} \cline{5-6}
 & \multicolumn{1}{c}{F555W} & \multicolumn{1}{c}{F814W} &
& \multicolumn{1}{c}{F555W} & \multicolumn{1}{c}{F814W} }
\startdata
22 -- 23&100.0&100.0&
&100.0&100.0 \\
23 -- 24& 97.0 $\pm$ \,~4.9& 99.9 $\pm$ \,~1.7&
&100.0&100.0 \\
24 -- 25& 91.7 $\pm$ \,~5.3& 96.1 $\pm$  \,~3.1&
&100.0& 98.2 $\pm$ \,~1.7 \\
25 -- 26& 90.6 $\pm$ \,~7.7& 89.7 $\pm$ \,~2.3&
& 97.4 $\pm$ \,~3.0& 98.2 $\pm$ \,~1.7 \\
26 -- 27& 85.1 $\pm$ \,~8.7& 85.7 $\pm$ \,~6.5&
& 95.4 $\pm$ \,~4.4& 94.5 $\pm$  \,~2.8 \\
27 -- 28& 78.9 $\pm$  \,~8.5& 70.7 $\pm$ 10.4&
& 92.9 $\pm$ \,~9.2& 80.4 $\pm$  \,~4.2 \\
28 -- 29& 55.3 $\pm$  15.7& 39.0 $\pm$ \,~6.3&
& 69.4 $\pm$ \,~9.9& 44.2 $\pm$ 10.1 \\
29 -- 30& 10.4 $\pm$ \,~7.7&  5.4 $\pm$ \,~7.1&
& 16.5 $\pm$ \,~9.4&  9.8 $\pm$ \,~8.4 \\
30 -- 31&  0.6 $\pm$ \,~2.5&  \nodata~~~~ &
& 0.6 $\pm$ \,~3.0& \nodata~~~~  \\
\enddata
\tablenotetext{a}{expressed in percentage of recovered stars.}
\end{deluxetable}

\clearpage

\clearpage

\clearpage

\clearpage

\clearpage

\clearpage

\clearpage

\clearpage

\clearpage

\clearpage

\clearpage

\clearpage

\clearpage

\end{document}